\documentclass[10pt, a4paper, iop,final]{emulateapj}
\def\jcap{J.\ Cosmo.\ \&\ Astro.}

\usepackage{lineno}
\usepackage{graphicx}
\usepackage{amssymb}
\usepackage{amsmath}
\usepackage{times}
\usepackage{courier}
\usepackage{xspace}
\newcommand{\fermi}{\emph{Fermi}}
\newcommand{\lat}{\emph{Fermi}-LAT}
\newcommand{\gr}{$\gamma$-ray\xspace}
\newcommand{\grs}{$\gamma$ rays\xspace}
\newcommand{\CR}{cosmic-ray}

\newcommand{\LL}{log-likelihood}
\newcommand{\is}{interstellar}
\newcommand{\iem}{IEM\xspace}
\newcommand{\iems}{IEMs\xspace}
\newcommand{\beq}{\begin{equation}}
\newcommand{\eeq}{\end{equation}}
\newcommand{\rmn}[1]{\ensuremath{\mathrm{#1}}}
\newcommand{\sigv}{\ensuremath{\langle\sigma v\rangle}\xspace}
\newcommand{\msol}{\ensuremath{M_{\odot}}}
\renewcommand{\deg}{\ensuremath{^{\circ}}\xspace}
\newcommand{\gev}{\ensuremath{\mathrm{GeV}}\xspace}
\newcommand{\tev}{\ensuremath{\mathrm{TeV}}\xspace}
\newcommand{\mev}{\ensuremath{\mathrm{MeV}}\xspace}
\newcommand{\xcr}{\ensuremath{\langle X_{\mathrm{CR}}\rangle}\xspace}
\newcommand{\zetap}{\ensuremath{\zeta_{p,\mathrm{max}}}\xspace}
\usepackage[usenames,dvipsnames]{color}

\usepackage[normalem]{ulem}

\newcommand{\edit}[1]{#1}
\newcommand{\medit}[1]{\ensuremath{#1}}

\begin{document}
\title{Search for extended gamma-ray emission from the Virgo galaxy cluster with \lat}
\shorttitle{Search for \grs from Virgo }
\shortauthors{Ackermann et al.}
\slugcomment{ApJ, accepted}
\author{
M.~Ackermann\altaffilmark{1}, 
M.~Ajello\altaffilmark{2}, 
A.~Albert\altaffilmark{3}, 
W.~B.~Atwood\altaffilmark{4}, 
L.~Baldini\altaffilmark{5,3}, 
G.~Barbiellini\altaffilmark{6,7}, 
D.~Bastieri\altaffilmark{8,9}, 
K.~Bechtol\altaffilmark{10}, 
R.~Bellazzini\altaffilmark{11}, 
E.~Bissaldi\altaffilmark{12}
% ...
\\
(The Fermi-LAT Collaboration)\\
and\\
A.~Pinzke\altaffilmark{53,29,28,63} 
}
  
\altaffiltext{1}{Deutsches Elektronen Synchrotron DESY, D-15738 Zeuthen, Germany}
\altaffiltext{2}{Department of Physics and Astronomy, Clemson University, Kinard Lab of Physics, Clemson, SC 29634-0978, USA}
\altaffiltext{3}{W. W. Hansen Experimental Physics Laboratory, Kavli Institute for Particle Astrophysics and Cosmology, Department of Physics and SLAC National Accelerator Laboratory, Stanford University, Stanford, CA 94305, USA}
\altaffiltext{4}{Santa Cruz Institute for Particle Physics, Department of Physics and Department of Astronomy and Astrophysics, University of California at Santa Cruz, Santa Cruz, CA 95064, USA}
\altaffiltext{5}{Universit\`a di Pisa and Istituto Nazionale di Fisica Nucleare, Sezione di Pisa I-56127 Pisa, Italy}
\altaffiltext{6}{Istituto Nazionale di Fisica Nucleare, Sezione di Trieste, I-34127 Trieste, Italy}
\altaffiltext{7}{Dipartimento di Fisica, Universit\`a di Trieste, I-34127 Trieste, Italy}
\altaffiltext{8}{Istituto Nazionale di Fisica Nucleare, Sezione di Padova, I-35131 Padova, Italy}
\altaffiltext{9}{Dipartimento di Fisica e Astronomia ``G. Galilei'', Universit\`a di Padova, I-35131 Padova, Italy}
\altaffiltext{10}{Dept.  of  Physics  and  Wisconsin  IceCube  Particle  Astrophysics  Center, University  of  Wisconsin, Madison,  WI  53706, USA}
\altaffiltext{11}{Istituto Nazionale di Fisica Nucleare, Sezione di Pisa, I-56127 Pisa, Italy}
\altaffiltext{12}{Istituto Nazionale di Fisica Nucleare, Sezione di Bari, I-70126 Bari, Italy}
% ...
\altaffiltext{28}{Department of Physics, Stockholm University, AlbaNova, SE-106 91 Stockholm, Sweden}
\altaffiltext{29}{The Oskar Klein Centre for Cosmoparticle Physics, AlbaNova, SE-106 91 Stockholm, Sweden}
% ...
\altaffiltext{53}{Dark Cosmology Centre, Niels Bohr Institute, University of Copenhagen, 2100 Copenhagen, Denmark}
% ...
\altaffiltext{63}{email: apinzke@dark-cosmology.dk}

\begin{abstract}
Galaxy clusters are one of the prime sites to search for dark matter (DM) annihilation signals. Depending on the substructure of the DM halo of a galaxy cluster and the cross sections for DM annihilation channels, these signals might be detectable by the latest generation of \gr\ telescopes. Here we use three years of \fermi-Large Area Telescope (LAT) data, which are the most suitable for searching for very extended emission in the vicinity of nearby Virgo galaxy cluster. Our analysis reveals statistically significant extended emission which can be well characterized by a uniformly emitting disk profile with a radius of 3\deg that moreover is offset from the cluster center. We demonstrate that the significance of this extended emission strongly depends on the adopted \is\ emission model (\iem) and is most likely an artifact of our incomplete description of the \iem in this region. We also search for and find new point source candidates in the region. We then derive conservative upper limits on the velocity-averaged DM pair annihilation cross section from Virgo. We take into account the potential \gr\ flux enhancement due to DM sub-halos and its complex morphology as a merging cluster. For DM annihilating into $b\overline{b}$, assuming a conservative sub-halo model setup, we find limits that are between 1 and 1.5 orders of magnitude above the expectation from the thermal cross section for $m_{\rmn{DM}}\lesssim100\,\gev$. In a more optimistic scenario, we exclude $\sigv\sim3\times10^{-26}\,\rmn{cm^{3}\,s^{-1}}$ for $m_{\rmn{DM}}\lesssim40\,\gev$ for the same channel. Finally, we derive upper limits on the \gr-flux produced by hadronic cosmic-ray interactions in the inter cluster medium. We find that the volume-averaged cosmic-ray-to-thermal pressure ratio is less than $\sim6\%$.
\end{abstract}

\section{Introduction}
Galaxy clusters are the largest gravitationally bound objects in the Universe. Observations of galaxy clusters, in particular, estimations of the gravitation potential based on measurements of the velocity dispersions of member galaxies, suggested that Galaxy cluster masses were much larger than the values inferred by summing contributions from luminous matter, and lead to the proposal of non-luminous (i.e., dark) matter (DM)~\citep{Zwicky1937}. Since then, many observations, as well as N-body cosmological simulations, suggest that galaxy clusters contain a high amount of DM, making them prime targets to search for indirect DM signals \citep[see, e.g.][]{Colafrancesco2006,Clowe:2006aa,2009PhRvD..80b3005J,2009PhRvL.103r1302P,MASC2011,Klypin2011,Pinzke2011,Gao2012a,Hellwing:2015aa}.

Weakly interacting massive particles (WIMPs) constitute promising particle DM candidates. Among prominent WIMP candidates is the neutralino, which in many supersymmetric models is the lightest stable supersymmetric particle, allowing it to account for the observed relic DM density in the Universe. In many of these models, the neutralino can self-annihilate into particle-antiparticle pairs which subsequently may produce other particles, including \grs. The \grs will propagate undeflected by the \is\ magnetic fields and thus reveal the location of the DM annihilation~\citep[see, e.g.][for a review of searches for indirect DM searches using \grs]{Bertone:2005aa,Feng:2010aa,Bringmann:2012aa,Conrad:2015aa}. Several predictions of the expected DM annihilation rate in cosmological environments, such as galaxy clusters, and the associated signals in \gr\ data show that current space-borne \gr\ detectors like the Large Area Telescope (LAT) on board the \fermi\ satellite \citep{LAT} will be unlikely to detect this signal in case the DM is smoothly distributed~\citep[see, e.g.][]{Pinzke2011}.

However, a smooth DM distribution is indeed not expected and recent cosmological N-body simulations predict instead that DM virialized regions, known as halos, contain a large number of smaller virialized and highly concentrated substructures called \emph{sub-halos} \citep{2005Natur.435..629S,2008Natur.454..735D,2008MNRAS.391.1685S}. Since the DM annihilation signal is proportional to the DM density squared, these highly concentrated sub-halos are expected to significantly boost the annihilation signal relative to the purely smooth DM scenarios \citep[e.g.,][]{2008ApJ...686..262K,2008MNRAS.391.1685S,2008A&A...479..427L,2008MNRAS.384.1627P,2009JCAP...06..014M}. The exact signal enhancement depends on the abundance, distribution and internal structural properties of the sub-halos. However, sub-halo properties are uncertain below the halo mass resolution of state-of-the-art N-body cosmological simulations, $\sim O(10^5\,\msol)$ for Milky Way size halos \citep{2008Natur.454..735D,2008Natur.456...73S,Hellwing:2015aa} and $10^8\,\msol$ for simulations of galaxy clusters \citep[e.g.,][]{Gao2012a}. Thus, extrapolations of the relevant properties are required over several orders of magnitude in halo mass below the mass resolution limit in order to account for the whole halo mass range that is predicted to exist in the Universe, and more specifically in clusters. Here, the fractional enhancement of \gr flux due to sub-halos is called the sub-halo boost, or boost-factor in short. As recently discussed e.g. in \citet{MASC2013}, sub-halo boosts are very sensitive to the way these extrapolations are performed and boost estimates can vary drastically depending on the assumptions \citep[e.g.,][]{2010PhRvD..81d3532K,MASC2011,Pinzke2011,Gao2012a,2012PDU.....1...50K,2012MNRAS.425..477N,2014MNRAS.441.1329Z}.
A debate is ongoing as to whether the extrapolation with a power law to lower-mass halos \citep{Pinzke2011,Gao2012a} is justified or too optimistic~\citep[see, e.g.][]{2008Natur.454..735D,MASC2011}.\footnote{Note that this inherent theoretical uncertainty is alleviated if decaying DM is considered \citep{Dugger:2010aa,Huang2012}.}

A further challenge arises from the fact that \CR\ (CR) interactions in the intra-cluster medium (ICM) may also give rise to $\mev--\gev$ \grs, which if observed, would be difficult to distinguish from a DM-induced signal \citep[see, e.g.][]{Pinzke2011,2012JCAP...07..017A}. Despite intensive efforts, to date no \grs\ from clusters have been detected aside from those attributed to individual active galaxies \citep[see, e.g.][]{2010ApJ...717L..71A,Huang2012,Huber2013,Ackermann2013}.

The closest galaxy cluster is Virgo at a distance of about $\medit{15.4\pm0.5\,\rmn{Mpc}}$, subtending several degrees on the sky~\citep{virgo_fouque}. The cluster consists of several sub-clusters which are located around giant elliptical galaxies, most prominently M87 and M49 as well as around the two smaller clusters associated with M100 and M60 \citep{2002astro.ph..6272S}. These sub-clusters are in the process of merging with one another, while the system is dominated by the most massive sub-cluster centered on M87. For the remainder of this work we refer to the sub-cluster centered on M87 as Virgo-I and the sub-cluster centered on M49 as Virgo-II. Also relevant for an analysis of \grs\ is the fact that M87 harbors a known active galactic nucleus \citep[AGN,][]{2009ApJ...707...55A,2014ApJ...788..165H} which dominates the emission both in X-rays and \grs. Due to its proximity, Virgo is also an interesting target for the search of a DM-induced \gr\ signal with the \emph{Fermi}-LAT.
Earlier studies, concentrating on Virgo-I, tested for point-like or mildly extended ($\leq 1\fdg2$) emission towards the center of the cluster and yielded upper limits on the integrated \gr\ flux of $14.1\times10^{-9}\mathrm{ph\,cm^{-2}\,s^{-1}}$ and $17.1\times10^{-9}\mathrm{ph\,cm^{-2}\,s^{-1}}$ respectively~\citep{2010ApJ...717L..71A}, assuming a power-law spectrum of the \gr\ emission with photon index $\Gamma=2$ in an energy band from 200~\mev to 100~\gev.

Claims of \gr\ emission induced by DM annihilation were put forward by \cite{HanI}, using a \emph{very extended} DM-induced emission profile considering only Virgo-I.\footnote{We use the term very extended to distinguish mildly extended emission (up to $\sim2--3\deg$ in radius) from larger extensions up to $\gtrsim7\deg$ in radius that are considered here. The former was tested in a recent work, yielding null results on extended \gr\, emission \citep{Ackermann2013}.} Later studies attribute this putative signal to an incomplete point source model of the region~\citep{Macias,HanII}.

Here we present a comprehensive analysis of the Virgo region searching for very extended emission and discuss the various systematic effects relevant for this analysis. In Section~\ref{sec:ana} we discuss our data selection and analysis. Sections~\ref{sec:virgoCenter} and \ref{sec:extended} elaborate on the details of finding and characterizing extended excess emission, while Section~\ref{sec:IEM} is devoted to the discussion of the uncertainties associated with the interstellar emission model (IEM). The aforementioned possibility of additional point sources constituting a putative signal is discussed in Section~\ref{sec:ps}. As a result of these studies we devise an improved background model for the Virgo region. We use this model to derive new limits on the WIMP DM annihilation cross section (Section~\ref{sec:dmVirgo}) and in the case of CR-induced \gr production on the its flux. We compare our results to theoretical predictions in order to constrain relevant CR quantities (Section~\ref{sec:crVirgo}). We conclude and summarize our work in Section~\ref{sec:conclusion}.

\section{Observations \& Data Analysis}\label{sec:ana}
The main instrument on board the \fermi\ satellite, the LAT, is a pair-conversion telescope sensitive to \grs in the energy range from 20~\mev\ to $>300~\gev$. For a more detailed description the reader is referred to \cite{LAT} and for the characterization of the on-board performance to \cite{fermi_inst2012}.

We analyzed archival \lat\ data between MJD 54682.7 (2008-08-04) and MJD 55789.5 (2011-08-16) corresponding to roughly three years of Pass~7 data. We chose Pass~7 data because the predicted \gr\ signal from possible DM annihilation stretches over several degrees in the case of the Virgo cluster. With the release of the reprocessed P7 data (P7REP), the LAT collaboration also released a new template to describe the Galactic foreground emission which is tuned to P7REP data. This model contains a component which is derived from re-injecting residual emission above a scale radius of about $2^{\circ}$\citep{Casandjian:2015aa}. Residuals larger than this scale radius are absorbed into the model and thus the model cannot be used to search for emission with larger extension. The release of P7REP data also implied a switch in the data processing pipeline limiting the available Pass~7 data to the three years which we used in this analysis. The reprocessing primarily impacts the energy reconstruction as it accounts for a time-dependent change of the calorimeter calibration constants \citep{ReprocessingPaper}, which in case of a signal may result in a moderate ($2--3\%$) shift towards higher photon energies with respect to Pass~7 data as well as an improvement in the high energy LAT point spread function (PSF), which however does not constitute any significant impact on our analysis.  {\edit{We also would like to point out that the recently released Pass~8 event selection provides  $\sim25$\% increased effective area above $1\,\gev$~\citep{Ackermann-M.:2015aa}~ and increases the data live time by a factor of two compared to the data set used in our analysis of the Virgo region.}} We will show in subsequent sections that our findings and interpretation are limited by systematic uncertainties in the interstellar emission model (IEM). \edit{These uncertainties can not be overcome by statistics but require better understanding of the CR distribution and its interaction in our Galaxy which is used to derive the IEM. Moreover, the IEM released with Pass~8 contains a data-driven residual component at the scale of 2 degrees. Hence, using this model is not suitable in the case of the spatially extended Virgo cluster.} 

The data were processed using the \emph{Fermi} {\tt{ScienceTools}} version v9r28p0. \footnote{The software packages required for LAT analysis along with the templates used to model the \is\ and extragalactic emission are made publicly available through the Fermi Science Support Center \url{http://fermi.gsfc.nasa.gov/ssc/data/}}
We selected events with high probability of being \grs by choosing the Source event class. In order to evade \gr\ contamination generated by CRs interacting with the atmosphere of the Earth, we removed events with a LAT zenith angle $>100\deg$. We excised time periods around bright solar flares and gamma-ray bursts and applied a rocking angle cut of 52\deg . Furthermore, we restricted our analysis to the 100~\mev\ to 100~\gev\ energy range and used the P7SOURCE\_V6 instrument response functions.

\begin{figure}[tbp]
  \centering
  \includegraphics[width=.9\columnwidth]{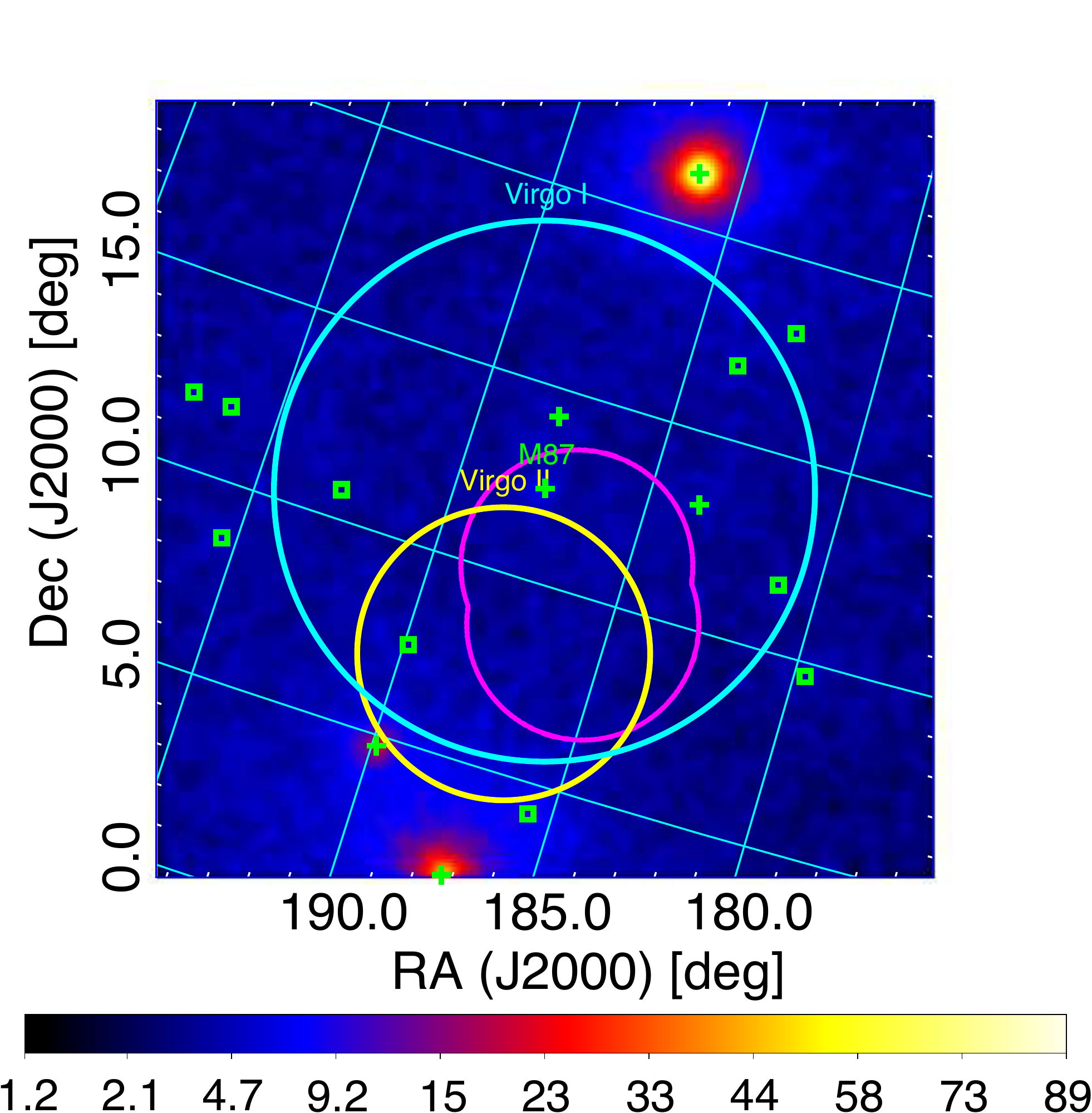}
  \caption{ Counts map of the Virgo ROI between $100\,\mev$ and $100\,\gev$ smoothed by a $\sigma=0\fdg3$ gaussian kernel. 2FGL sources with free parameters in the likelihood fit are marked by crosses and those that have fixed parameters are marked by boxes. The prominent AGN M87 is almost in the cluster center. The cyan and yellow circles correspond to the angle subtending the virial radius, of Virgo-I and Virgo-II, respectively (see Section~\ref{sec:dmVirgo} for details). We show the excess identified in Section~\ref{sec:extended} as a magenta contour.}
  \label{fig:cmap}
\end{figure}

To model the Galactic foreground emission caused by CRs interacting with the gas and radiation fields in our Galaxy, we use the {\tt{gll\_iem\_v02.fit}} model. This IEM is the standard \iem provided by the \lat\ collaboration for point source analysis in the un-reprocessed flavor. The isotropic \gr\ emission is accounted for by the {\tt{isotropic\_iem\_v02.txt}} model. For simplicity we refer to this set of \iem and isotropic model as our \emph{standard} model. We chose a $20\deg\times20\deg$ region of interest (ROI) centered on the center of Virgo-I ($\alpha_{2000}=187\fdg71$ and $\delta_{2000}=12\fdg39$) and performed a binned likelihood analysis with $0\fdg1$ spatial bins and 30 logarithmic bins in energy.\footnote{The coordinates for Virgo-II are taken to be $\alpha_{2000}=187\fdg45$ and $\delta_{2000}=8\fdg00$.} A counts map of the ROI together with the position of all sources from the LAT 2-year catalog~\citep{2FGL} sources is shown in Fig.~\ref{fig:cmap}. In addition to the two aforementioned diffuse model components, our background model contains all sources within a $30^{\circ}$ radius around the Virgo-I center that are listed in the 2FGL catalog~\citep{2FGL}. The Virgo ROI contains mostly extragalactic sources which may be variable and thus the two-year source parameters in the 2FGL catalog might be bad approximations for the three-year data. Another challenge arises by performing an analysis down to 100 MeV. At this energy the \lat\ PSF with 68\% containment radius is about  $7^{\circ}$~\citep{fermi_inst2012} and thus even far away but strong sources which are not modeled correctly might easily increase the significance of a very extended profile located at the cluster center. To account for this we free the normalization and spectral index of all sources within 5\deg from the center coordinates of either Virgo-I or II in addition to the bright sources in the ROI (see below). The sources left free in our fit are marked by crosses in Fig.~\ref{fig:cmap}.

Using a likelihood analysis we construct a test statistic ($TS$) following~\citet{ts} to evaluate the improvement of the likelihood fit to the ROI when adding a new source by defining $TS=-2(\log\mathcal{L}_{0} - \log\mathcal{L})$. $\mathcal{L}_{0}$ refers to the maximum likelihood value for the null hypothesis and $\mathcal{L}$ for the value for the alternative hypothesis (including an additional source such as the cluster itself). In the case of one additional degree of freedom the significance can be written as $\sigma=\sqrt{TS}$ (see Section~\ref{sec:extended} for details regarding the gauging of this quantity). We look for new point-like excesses and derive $TS$-maps by placing a test point source at the location of each pixel of the map and maximizing the likelihood using the {\tt{gttsmap}} tool.

To estimate the systematic uncertainties on source significance and flux properties caused by our limited knowledge of the \iem, we compare the standard \iem results to the results obtained with eight alternative models seeded with \gr emission maps generated by GALPROP\footnote{\url{http://galprop.stanford.edu}}~\citep{Vladimirov:2011aa} and additional templates (as described in detail in Section~\ref{sec:IEM}) \citep{2013arXiv1304.1395D}. We would like to stress that these models do not cover the entire systematic uncertainty associated with the IEM and the results are not expected to bracket the standard IEM results. Despite the potentially small coverage, using these models demonstrates the influence the IEM has on our result.

\section{Search towards the Virgo Cluster Center}\label{sec:virgoCenter}

\begin{figure}[tbp]
  \centering
  \includegraphics[width=.9\columnwidth]{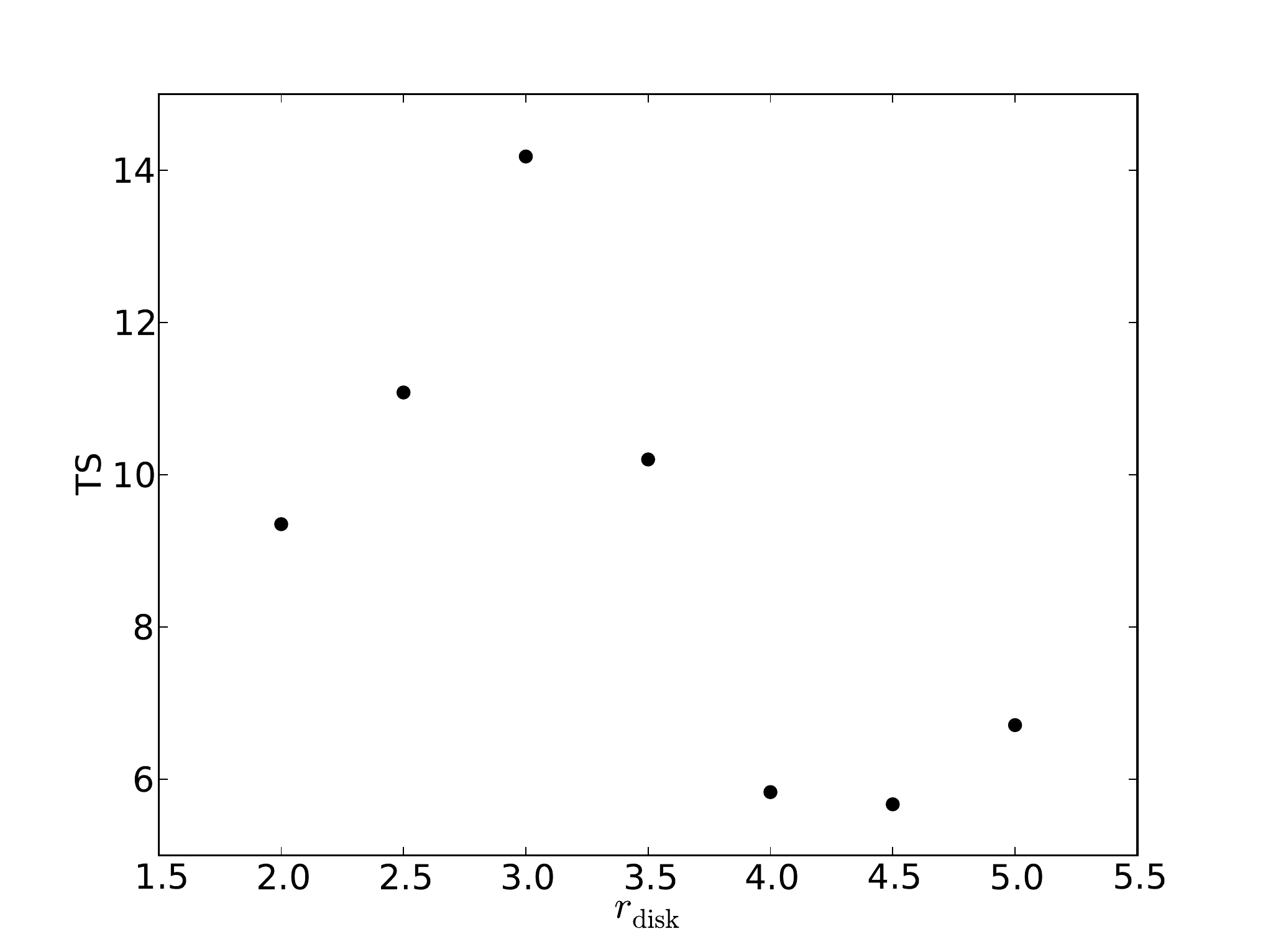}
  \caption{$TS$ value for adding an extended source with a uniform disk profile fixed at the center of Virgo~I vs the radius of the disk ($r_{\textrm{disk}}$). There is a clear peak in the $TS$ distribution (at $r_{\textrm{disk}}=3\deg$) as it is expected from a finite-sized source of excess \gr . In case of an excess due to an overall residual photon distribution in the whole ROI a steady increase would be expected, which is incompatible with our findings.}
  \label{fig:disk_r}
\end{figure}

Instead of starting our search for extended emission from the cluster center with a specific physical model, we consider a more generic model for any potential excess, namely a simple uniform disk with a given radius, $r_{\mathrm{disk}}$, centered on Virgo~I. The uniform disk profile is very successful in finding weak extended emission and is less prone to degeneracy with strong point sources such as M87. Moreover, a disk profile is usually sufficient to find sources of various shapes because, even for very strong sources, a discrimination between different emission profiles is usually not possible \citep[see, e.g.][]{fermi_extended}. For these reasons, we use the disk profile with a power-law spectral model for all our systematic studies of the extended excess (Sections~\ref{sec:virgoCenter} and \ref{sec:extended}). To further characterize the extended excess, we perform a $TS$ vs radius scan. The scan is performed in $0\fdg5$ steps of $r_{\mathrm{disk}}$ and shows a peak at $r_{\mathrm{disk}}=3\deg$ with an associated $TS$-value of 14.2 as shown in Fig.~\ref{fig:disk_r}. This finding is in agreement with \citet{HanI} who attributes most of their found emission to the innermost $3\deg$ of their profile.

Next, we considered the case of a DM-induced signal as proposed by~\citet{HanI} which would lead to a \gr\ contribution peaked at the cluster center, again considering Virgo-I as the center of the cluster. To this end we substitute the disk profile with the DM profile used in \citet{HanI} in our source model of the Virgo region. We reproduce their results, finding $TS\simeq23$, only if we fix the bright sources outside of $5\deg$ of Virgo-I to their catalog values as reported in 2FGL. However, when performing the analysis according to the description in Section~\ref{sec:ana}, the $TS$-value drops considerably ($TS\simeq17$).

\section{Origin of the putative extended emission}\label{sec:extended}

\begin{figure}[tbp]
  \centering
  \includegraphics[width=.9\columnwidth]{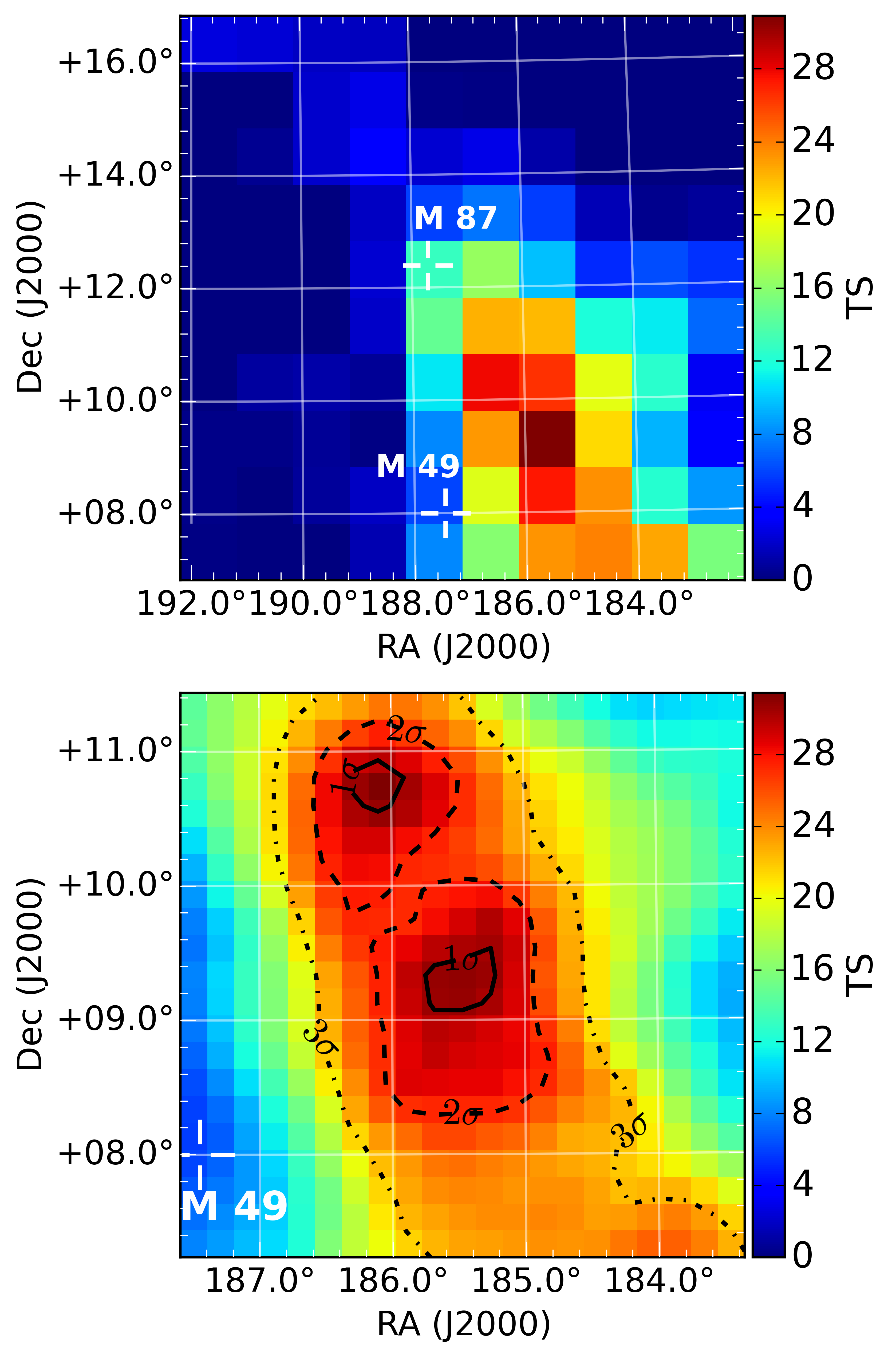}
  \caption{ \emph{Top:}  $TS$-map of a uniform disk with $r_{\mathrm{disk}}=3\deg$ using a 1\fdg0 grid for the $10\deg\times10\deg$ Virgo region. The open crosses mark the center position of Virgo-I and II. There is a concentration of high $TS$ values in the lower right quadrant indicating that the position of the centroid is not well defined. \emph{Bottom:}  $4\deg\times4\deg$ region of the coarser $TS$-map shown in the top panel. The finer 0\fdg2 binning emphasizes two broad $TS$ maxima of equal height which are well separated by about 1\fdg5. The different contours indicate the 1, 2, and $3\sigma$ levels.}
  \label{fig:tsmap_ext}
\end{figure}

We scan the inner $10\deg\times10\deg$ of the Virgo region with a disk with a radius of $r_{\mathrm{disk}}=3\deg$ on a 0\fdg5 grid to find the best position for the origin of the putative extended excess. To obtain this $TS$ map we add a disk-like test source at each grid position and maximize the likelihood using {\tt{gtlike}}. We leave all sources within a radius of $5\deg$ from the position of the test source and all bright sources ($TS > 1000$)  free to vary in normalization and spectral index. This approach is needed since in the presence of a real extended source, the degeneracy between the extended source and overlapping or very close-by point sources will decrease the significance of the extended sources. The resulting $TS$-map has peaks significantly offset from the centers of Virgo-I and Virgo-II. All of them are located in the lower right quadrant of our $TS$ map shown in Fig.~\ref{fig:tsmap_ext}.
To obtain a more detailed description we made a finer grid of 0\fdg2  spacing in the $4\deg\times4\deg$ region encompassing the highest $TS$ values. This finely binned $TS$-map is shown together with the coarse map in Fig.~\ref{fig:tsmap_ext}. From the fine map it is evident that a large region of the Virgo ROI yields $TS$ values above 25. In particular, there are two broad maxima seen which are spatially distinct from one another. Note that the typical $1\sigma$ localization contour for a source near threshold is about $0.1\deg$. Consequently, the peaks shown in Fig.~\ref{fig:tsmap_ext} appear to be larger than what would be expected from a point source.
{\edit{Repeating the study of $r_\rmn{disk}$ vs. TS at the two maxima positions we again find a clear peak at $r_{\mathrm{disk}}=3\deg$ and thus continue all subsequent analysis using a disk with $r_{\mathrm{disk}}=3\deg$. }}

\begin{figure}[tbp]
  \centering
  \includegraphics[width=.9\columnwidth]{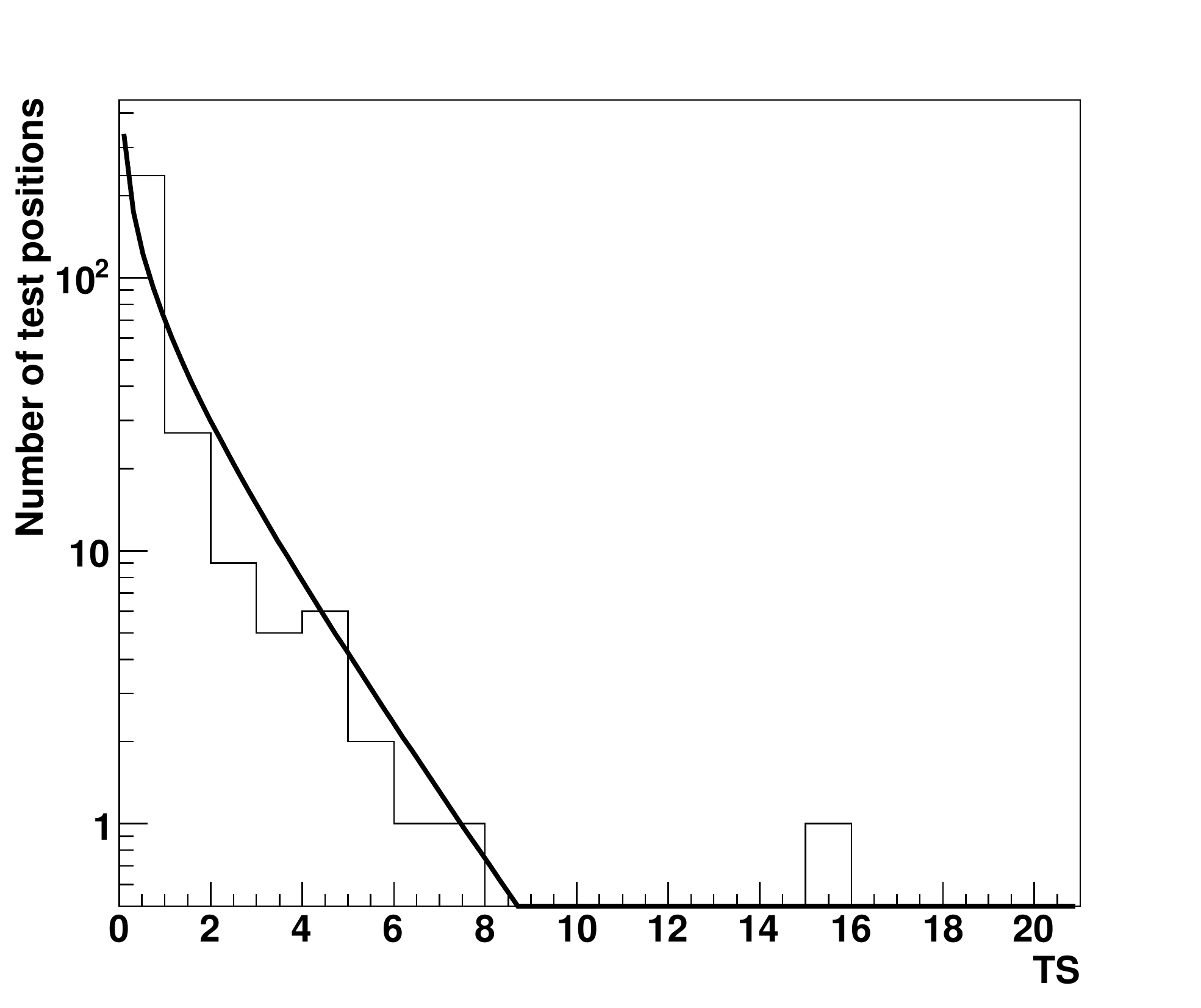}
  \caption{$TS$ distribution for adding an extended source with a uniform disk profile $r_{\textrm{disk}}=3\deg$ at randomly selected high Galactic latitude ($|b|>20\deg$) positions. The ROIs of the positions do not overlap within their inner $5\deg$. We do not find any $TS$ above 16 and most of the values lie below eight. The solid line is a $\chi^2$ distribution with one degree of freedom that reasonably well describes the data.}
  \label{fig:ts_hist}
\end{figure}

In order to study the relationship between $TS$-values and significance, we performed an extended source search at 288 randomly selected sky positions (blank fields) to estimate the significance of finding a disk-shaped excess with $r_{\mathrm{disk}}=3\deg$. All test positions were selected so that their inner $5\deg$ do not overlap and hence each position can be treated as statistically independent. Furthermore, we only used positions with $|b|>20\degr$ and therefore exclude regions with bright Galactic \gr\ emission. The resulting TS distribution is shown in Fig.~\ref{fig:ts_hist} and is reasonably described by a $\chi^2$ distribution with one degree of freedom. We note that the number of tested blank fields is too small to sample any probability density function at $TS>8$. However, the absence of $TS>16$ along with the fact that the majority of blank fields yield $TS<8$ indicate that the significance of the excess we find here  (up to $TS \sim 32$) is larger than $3\,\sigma$. We stress that this particular statement is only valid at high Galactic latitude from where we extracted our blank fields.

\section{Interstellar Emission}\label{sec:IEM}
\begin{figure}[tbp]
  \centering
  \includegraphics[width=.9\columnwidth]{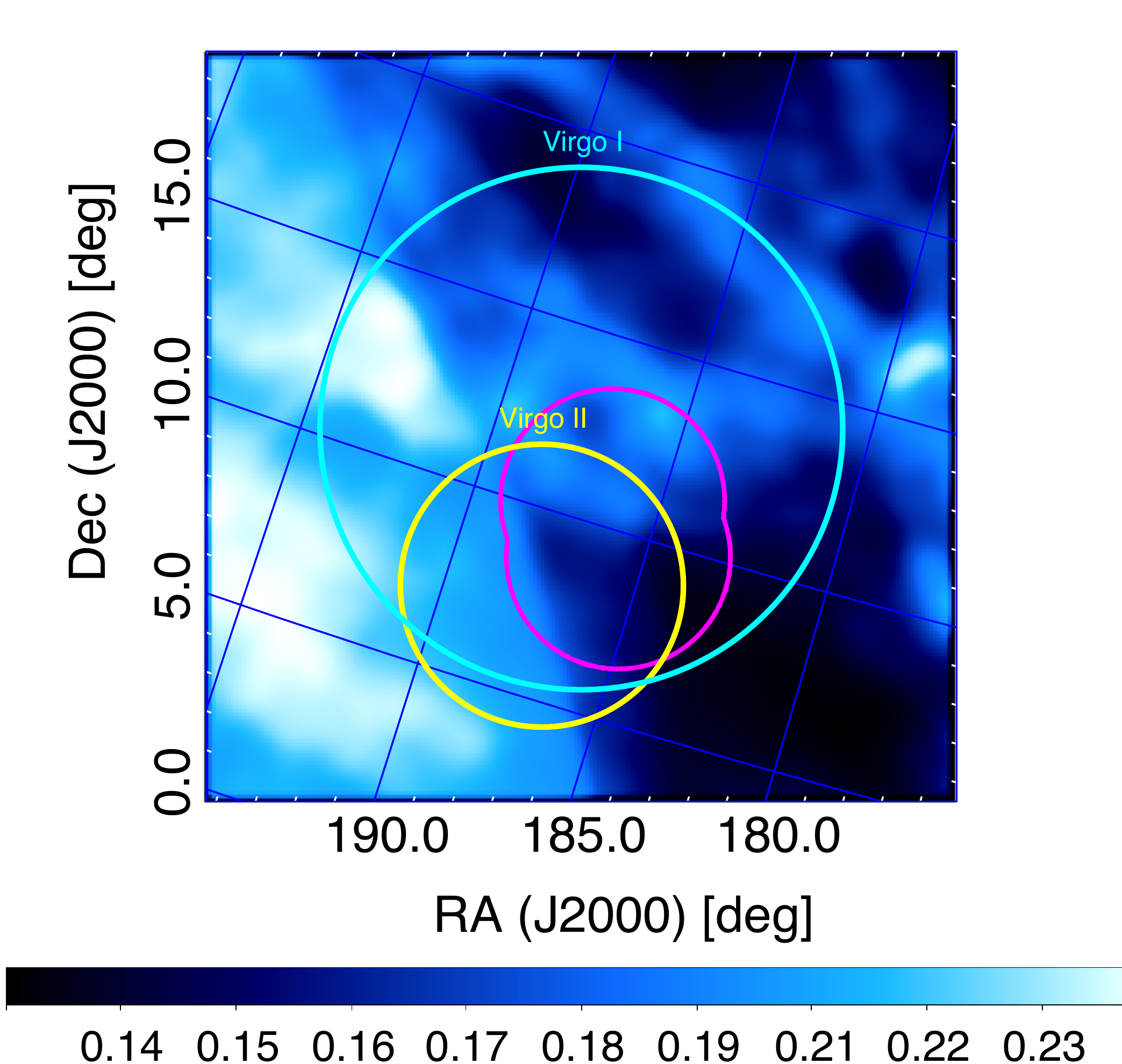}
  \caption{Model counts map of the standard \iem model for the Virgo ROI above $1\,\gev$. The cyan and yellow circles correspond to the angle subtending the virial radius, of Virgo-I and Virgo-II, respectively (see Section~\ref{sec:dmVirgo} for details). We show the excess identified in Section~\ref{sec:extended} as a magenta contour. Clearly visible is the patch that we associated with Loop~I as the bright light-blue band on the left side and several bright filaments, especially a donut-like shape close to the center of the ROI. The Virgo region does not show a uniform \iem as might be expected for high-latitude ROIs and must be treated accordingly when searching for extended emission.}
  \label{fig:iem}
\end{figure}

The interstellar \gr\ emission is the dominant \gr\ background in \lat\ analyses in the vicinity of the Galactic plane, while far from the Galactic plane the largest \gr\ contribution usually stems from discrete sources and isotropic \gr\ emission. However, in certain high-latitude regions the interstellar emission can contribute significantly to the \gr\ background and one of these regions is Virgo. Several features present in the \iem are contained within the Virgo ROI. The most striking features are the spatially uniform patch that is used to model the \gr\ emission from Loop~I~\citep{1962MNRAS.124..405L} and some filaments of HI around the center of the Virgo cluster as shown in Fig~\ref{fig:iem}.\footnote{See \url{http://fermi.gsfc.nasa.gov/ssc/data/access/lat/Model\_details/Pass7\_galactic.html} for the description of the model of Loop~I.}
Loop I is only approximately modeled in the standard \iem and could influence our analysis of the region. For most point sources in the Virgo region the \iem does not play a significant role since it contributes on average only about 110 counts per square degree above $100\,\mev$. Yet, it must be considered carefully when analyzing weak and very extended emissions like our disk model in the previous section (accounting for about 800 predicted photons and about 28 counts per square degree). The isotropic \gr\ emission has the largest contribution to the diffuse \gr\ emission in the ROI and accounts for about 200 counts per square degree. We note that the spatial variation in the \iem \gr\ counts are up to a factor of two from the minimum value. These spatial variations in the IEM are about three times larger than the potential \gr\ counts from the disk profile. The weak disk emission could easily be caused by a missing feature in the \iem . Such a missing component  would necessarily not be traceable by tracers of the interstellar gas (21-cm H\,{\sc I} line and 2.6-mm CO line), and makes the systematic evaluation of \iem\ influences on our analysis mandatory.

To assess the influence of the uncertainty in our knowledge of the \iem, besides the standard \iem, we use eight additional models. These models are seeded with simulated \gr\ intensities obtained from GALPROP, assuming CR halo heights of 4~kpc and 10~kpc, two different CR source populations, referred to as 'pulsars' and 'supernova remnants' (SNR) and two different HI spin temperatures (optically thin and 150K). Afterwards each model's spectral component is multiplied with a log-parabola function to provide a better model-data agreement and thus is not a direct output of GALPROP. For further details on the alternative models see~\cite{2013arXiv1304.1395D}.  Most of the parameters varied in the alternative models are only expected to have very slight influence on the results for the Virgo region since they should not influence the local CR density very much and that is what is mostly sampled when looking in the direction of Virgo.
For each of these alternative models an individual modified isotropic diffuse contribution is used.

Each of these models is comprised of maps that are inferred from HI and CO tracers along with components modeling the large scale diffuse residual structures, such as Loop~I~\citep{fermi_loopI} and the \fermi\ bubbles~\citep{fermi_bubbles2010,Bubbles_2014}. The HII emission that is inferred from GALPROP and based on the NE2001 model is added to the HI map. In addition each model also includes a model of emission from inverse Compton (IC) scattering of CR electrons on interstellar radiation also calculated by GALPROP.  For these models the spectral line shifts of the HI and CO lines were used to derive maps for separate ranges of Galactocentric distance. Inside the Virgo ROI only HI is present and is located  within the second and third ring HI templates (these rings correspond to Galactocentric distance ranges 4--8 kpc and 8--10 kpc, respectively). Beside these two HI components only the IC model and the Loop~I template are included in our fit. Each normalization of the alternative \iems components is left free in the likelihood fit. The second HI ring only contributes to the Virgo ROI because of the HII emission that is added to the HI template. We note that there is considerable uncertainty in the estimation of the HII emission.  The Loop~I template is a geometrical template adapted from~\cite{2007ApJ...664..349W} and based on a polarization survey at 1.4 GHz and it is modeled as two expanding shells centered on two local OB associations.
Our standard IEM on the other hand uses a uniform-patch Loop~I template whose shape was derived by visual inspection of the gamma-ray residuals when building the standard IEM\footnote{There is also another component in the Loop~I template of the standard IEM derived from from 408 MHz radio maps\
  \citep{1982A&AS...47....1H} but this template is not visible in the Virgo ROI.}. There is currently no template of Loop~I available from observations at other wavelengths that adequately traces the gamma-ray emission observed by Fermi-LAT in the direction of Loop~I. By following two different approaches to define a Loop~I template into our models we can gain some insights into the influence of the Loop~I modeling on our results.

We randomly select one position from the bottom panel of Fig.~\ref{fig:tsmap_ext} where $TS\geq25$ and repeat our likelihood calculations using these alternative diffuse models.

The results do not have consistent $TS$ values between the individual \iems as can be seen in Table~\ref{tab:iem}. While the standard, I, V and VII \iems show high $TS$ values for an extended source, the other models yield considerably lower values.
Leaving the individual components of the \iem free to vary in the fit allows for a higher sensitivity to features in the \iem which might affect only one component map. Such a mis-modeling of a single component in the \iem could cause an extended excess roughly corresponding in shape with the mis-modeled or missing emission.

The large observed variation in $TS$-values of the disk emission for the alternative models provides an indication that the observed \gr\ emission may be due at least partially to an incomplete modeling of the \iem. In the direction of the Virgo cluster the interstellar gas is mostly local (within $\sim$1~kpc of the Sun) and thus only the local CRs should contribute to the \iem. The \gr\ emission caused by the local CR density is not expected to have large dependencies on the CR source distribution, CR halo height, or spin temperature and thus we would expect rather similar results for all models. However the alternative \iems that have a relatively large $TS$ for the disk are associated with large predicted photon counts, overemphasising the contribution from HI ring 2 by increasing its amplitude by a factor O(10--30). Such an implausibly large increase in the contribution is explained by the shape of the model components of each \iem. HI ring 2 covers a region in the projected sky that is similar to the Loop I template in the standard model and by upscaling the normalization beyond the physically viable bounds of the HI component the overall fit is improved. Note that the extension of HI ring 2 in the ROI depends on the CR halo height and thus introduces some dependency of the $TS$ of the disk on this \iem parameter that is also not expected from the local CR density.  The high normalization of the HI ring 2  demonstrates that some \gr\ emission is coming from its large region within the ROI that is not traced by the usual HI and CO tracers but can be partially compensated by overestimating their contribution in the ROI. The disk emission might just compensate a part of this large-area diffuse gamma-ray emission that is not overlapped by any other \iem component that could compensate it.

In general it is very difficult to obtain a conclusive picture as to the possible emission in the ROI. The \iem study suggests that while there may be some additional emission, the ROI contains several \iem components whose predicted emission overlap with the excess, making it difficult to disentangle their contribution in a \LL\ fit to the \gr-data. A precise modeling of Loop I on the other hand, poses a considerable challenge since its \gr\ emission is not well traced by the radio emission. Hence, we used a geometric model for Loop I in the alternative \iems compared to the \gr\ residual-inferred template in the standard \iem. The low significance of the emission found in this work makes it extremely difficult to identify its origin due to the aforementioned issues but is likely caused by inaccuracies in the \iem .
To describe the location of the extended emission we devised a double disk patch that accounts for the observed residual \gr\ emission. This double disk consists of two disks with $r_{\mathrm{disk}}=3^{\circ}$, one at each of the maxima in our $TS$-map shown in Fig.~\ref{fig:tsmap_ext} and a uniform single power law emission of the whole profile. In this way the model covers most of the extended emission and we can easily trace its position in sky maps and include it in all relevant plots discussing possible counter parts. In all further analysis when we are deriving upper limits on \gr emission from DM annihilation or CR interaction we leave the newly found extended emission unmodeled to be conservative as we can not make a definitive statement about the origin of the extended emission. This results in weaker upper limits compared to when including the aforementioned double-disk model for the emission when deriving upper limits.

\begin{deluxetable*}{lcccccccc}
  \tablewidth{0pt}
  \tablecaption{Parameters for the 9 \is\ emission models used}
  \tabletypesize{\small}
  \vspace*{-0.3cm}
  \startdata

  \hline\hline
  Model    & Sources & Halo height & $T_{S}$ & $\log(L_0)$ & $\log(L_{\mathrm{disk}})$ &$TS_{\mathrm{disk}}$  & $F(E>1\,\gev)$ & $\Gamma$\\
  & & kpc   &        K  &  & &  & $\times10^{-9}\textrm{cm}^{-2}\textrm{s}^{-1}$&\\
  \hline
  I & Pulsars & 10 & $10^5$ & -342483.4   &-342472.1&21.7&$1.5\pm0.1$&$1.85\pm0.02$\\
  II & Pulsars & 4 & $10^5$ & -342484.0  & -342476.3 & 15.4 &$1.3\pm0.3$&$1.8\pm0.1$\\
  III & Pulsars & 10 & $150$ & -342480.3  & -342474.0  & 12.6&$1.3\pm0.3$&$1.8\pm0.1$ \\
  IV & Pulsars & 4 & $150$ & -342481.3  & -342474.0  & 14.6 &$1.3\pm0.3$&$1.8\pm0.1$ \\
  V & SNR & 10 & $10^5$ &-342481.9 &-342471.3 & 21.2&$1.5\pm0.4$&$1.9\pm0.1$\\
  VI & SNR & 4 & $10^5$ & -342484.0 &-342476.3 & 15.4&$1.3\pm0.3$&$1.8\pm0.1$\\
  VII & SNR & 10 & $150$& -342479.2  & -342469.4  & 19.7&$1.5\pm0.3$&$1.8\pm0.1$\\
  VIII & SNR & 4 &$150$& -342481.1  &-342474.1 & 14.1&$1.2\pm0.4$&$1.8\pm0.1$\\
  \hline
  Standard & \nodata & \nodata & \nodata & -342494.3  & -342480.4 & 27.4&$1.7\pm0.3$&$1.9\pm0.1$ \\
  \hline
  \enddata
  \tablecomments{The models I-VIII have the normalization of each individual component left free in the fit while the standard model has fixed ratios between individual components. The columns contain from left to right, the model name, the source population causing the \gr\ emission, the CR  halo height, the spin temperature, the reference \LL , the \LL\ with the added disk, the \gr\ flux and spectral index of the disk profile.\label{tab:iem}}
\end{deluxetable*}

\section{Point source search}\label{sec:ps}
Previously undetected (variable) point sources that are not contained in the 2FGL catalog provide a possibility to account for the extended excess we find. Such sources can be significant background sources in the three-year data set and should be included in the background model. This possibility was first addressed by~\cite{Macias} and later by~\cite{HanII}.
However, neither searched for possible extended emission (not from DM annihilation) offset from the cluster center. Therefore we performed our own point source search in the Virgo ROI to be able to compare these results with the same analysis set up as our extended source search. We note that we search for new point source candidates without the double disk profile described in the previous section. This allows us to construct a new alternative background model including the newly found point source candidates but not considering any additional extended source profile.

We divide the Virgo ROI in $0\fdg1\times0\fdg1$ grid positions and fit all 2FGL sources in the region to obtain the log-likelihood reference value. We then add one point source at one grid point and calculate the $TS$ of this source candidate and repeat this procedure at each grid point. We identify source candidates corresponding to grid points with $TS>15$ and use them as seeds for the \texttt{gtfindsrc} localization algorithm. Finally we add all source candidates into our Virgo ROI model and fit the region again to obtain a better reference model for the extended source investigation.

\begin{figure}[tbp]
  \centering
  \includegraphics[width=.9\columnwidth]{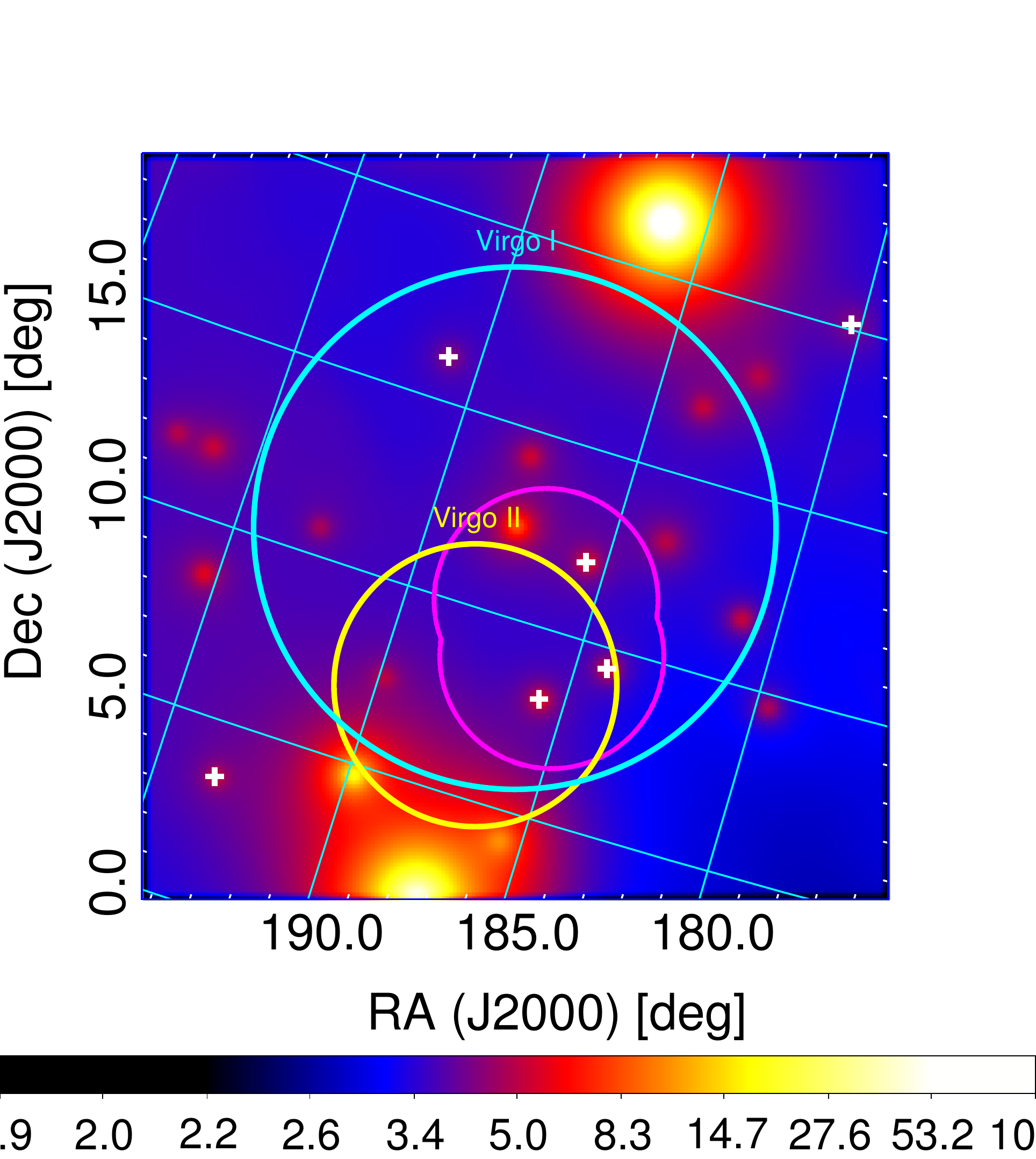}
  \caption{Model counts map generated with the best-fit parameters after adding six new point-source candidates to the Virgo ROI background model, integrated over the entire energy range ($100\,\mev \leq E \leq 100\,\gev$). The cyan and yellow circles correspond to the angle subtending the virial radius, of Virgo-I and Virgo-II, respectively (see Section~\ref{sec:dmVirgo} for details). We show the excess identified in Section~\ref{sec:extended} as a magenta contour.  The new sources are marked by crosses (white).}
  \label{fig:ps_model}
\end{figure}

Following this procedure, we find eight new source candidates, three of which appear clustered in close vicinity to one another. We note that when including either of the three, the remaining candidates are not statistically significant. Thus we only keep the brightest candidate in our background model and discard the other two. In summary, our new background model contains six new source candidates listed in Table~\ref{tab:ps}. Four of the six sources lie in the vicinity of the center of Virgo-I and three of them are in the region where we find the extended emission (substantially offset from the center of Virgo-I). While adding these six source candidates to our model yields a much better description of the ROI, the closest candidate sources to the center of Virgo-I or Virgo-II are below the conventional source detection threshold  ($TS>25$). Note that five out of these six sources are contained in the latest 4-year source catalog \citep[3FGL, see][]{3fgl}. To identify possible multi-wavelength counterparts to the \gr\ sources we searched in the 95\% error circle around each source in the  NASA/IPAC Extragalactic Database (NED) for potential \gr\ emitters. We do not find convincing counterparts for the new point source candidates; however, that does not perclude the possibility that they are real sources since a considerable fraction of \lat\ sources are not associated with multi-wavelength counterparts \citep{2FGL,3fgl}. The locations of the new source candidates are shown in the model map\footnote{A model map is the predicted counts map calculated for the ROI from the maximum-likelihood values of the model parameters.} in Fig.~\ref{fig:ps_model}. While we do get slightly different results compared to the work of~\cite{Macias} and~\cite{HanII} the discrepancies can be attributed to the larger data set used in~\cite{Macias} and~\cite{HanII} and the slightly different analysis procedures, such as the number of source parameters freed in the search or the different definitions of the ROIs. Considering the very weak emission of all source candidates the consistency of the findings among these works is reasonable. {\edit{We also note that when including the disk profile in addition to the new point source candidates in the ROI we obtain a $TS$ value of only $\sim7.5$ for the disk emission.}}

With this improved background model in hand, {\edit{which contains new point source candidates but not the disk profile}}, we devote the remainder of this paper to study the Virgo cluster as a \gr emitter, either via DM annihilation or via CR interactions (Sections~\ref{sec:dmVirgo} and~\ref{sec:crVirgo}).

\begin{deluxetable}{cccccc}
  \tablewidth{0pt}
  \tablecaption{Point source candidates not included in the 2FGL catalog}

  \vspace*{-0.3cm}
  \startdata
  \hline\hline
  Source candidate   & RA & DEC & 95\% error & $TS$ &3FGL source \\
  \hline
  I & 185.493 & 12.03&0.1& 23.5& J1223.2+1215\\
  II & 184.167& 9.460 &0.1&  23.4 & \\
  III &185.692 & 8.148 &0.1& 21.6 & J1223.3+0818\\
  IV & 190.89& 16.183&0.07& 28.1& J1244.1+1615\\
  V & 193.419 & 3.574&0.05& 49.6& J1253.7+0327\\
  VI & 180.292 & 20.141 &0.09& 27.8& J1200.9+2010\\
  \hline
  \enddata
  \label{tab:ps}
  \tablecomments{Note that 3FGL employs four years of data and typically contains sources with $TS>25$.}
\end{deluxetable}

\section{Search for Dark Matter Annihilation in Virgo}\label{sec:dmVirgo}
The $\gamma$-ray flux from annihilating DM particles of mass $m_{\chi}$ can be written as
\beq
\Phi_{\gamma}(E,\psi)=\frac{1}{4\pi}\frac{\sigv}{2 m_{\chi}} \sum_{j}{B_{j}\frac{\rmn{d}N_{j}(E)}{\rmn{d}E_{j}}} \times J(\psi).
\eeq

In the above equation, we have \sigv as the thermally-averaged product of DM self-annihilation cross section times velocity, and the sum runs over the final states the DM particles annihilate into with their specific \gr annihilation yields $\rmn{d}N_{j}(E)/\rmn{d}E_{j}$ and branching fraction $B_{j}$ per final state $j$. We define the astrophysical J-factor as the line-of-sight integral of the squared DM density towards the observational direction, $\psi$, integrated over a solid angle $\Delta\Omega$:\beq J(\psi)=\int_{\Delta\Omega}\left\{\int_{l.o.s.}{\rho^{2}{[l(\psi)]}}\mathrm{d}l\right\}\mathrm{d}\Omega'. \label{eq:los}\eeq
Density profiles $\rho(r)$ for plausible DM distributions can be expressed in terms of a generalized Hernquist profile \citep{Hernquist1990}:

\beq
\rho(r)=\rho_0\left(\frac{r}{r_s}\right)^{-\gamma}\left[1+\left(\frac{r}{r_s}\right)^{\alpha}\right]^{\frac{\gamma-\beta}{\alpha}},
\label{eq:nfw}
\eeq

where $\alpha,\beta$ and $\gamma$ are shape parameters. High-resolution cosmological N-body simulations of cold DM halos indicate that their density profiles are well described by a Navarro-Frenk-White (NFW) profile where $\alpha=1,\beta=3,\gamma=1$ \citep{Navarro1996,Navarro1997}. The quantity $r_s$ is the characteristic scale radius of the profile with $c$ as the concentration parameter such that $r_s = r_{200}/ c$. $r_{200}$ is defined to be the virial radius for which within the cluster mass $M_{200}$ is:
{\edit{
    $M_{200}= 4 \pi \rho_c/3 \times 200 \times r_{200}^3,$
    where $\rho_c$ is the critical density of the universe and $\rho_0$ is the characteristic density of the profile, $\rho_0= \frac{200}{3} \frac{c^3 \rho_c}{\ln{(1+c)}-c/(1+c)}.$

    As for the concentration-mass relation, we adopt here the one proposed by \citet{Prada2012} for the WMAP5 cosmology. This relation was derived from the $\sim10$ billion particle Bolshoi and MultiDark large scale structure N-body cosmological simulations \citep{Klypin2011}.\footnote{We assume a $\Lambda$CDM cosmology, characterized through $\Omega_{m}=0.32$, $\Omega_{\Lambda}=0.68$ and $h=0.67$ \citep{PlanckCollaboration2013}} By evaluating the integral $M_{500}=\int_{0}^{r_{500}}{ \rmn{d}^{3}r\,\rho(r)}$ with $\rho(r)$ as defined in Eq.~\eqref{eq:nfw}, we can numerically determine the value for $M_{200}$ for which this equation is satisfied, using the reported values for $M_{500}$ and $r_{500}$ from \citet{chen2007}.}}
We find $M_{200}=5.6\times10^{14}\,\msol$ as the mass of Virgo-I with $c=4.21$.

The integrated $J$-factor, for $r<r_{200}$ and assuming an NFW DM density profile, at a large angular diameter distance $D_a$ from Earth to the center of the cluster can be approximated by:
\beq J^{\rm NFW} = \frac{4\pi}3 \rho_0^2 r_s^3 \frac{1}{{D_a}^2}.\eeq

\subsection{Contribution from Substructure\label{sec:DMsubstructure}}
Some fraction of the DM should reside in sub-halos within the NFW-like primary halo. The presence of sub-halos implies both a flattening of the surface brightness profile \citep[see, e.g.][]{MASC2011,Gao2012a} and an enhancement (\emph{boost}, denoted $b$)  of the J-factor which may increase the total annihilation signal by orders of magnitudes. Here $b=0$ corresponds to the case of the smooth NFW halo without the inclusion of additional substructure.

For the normalization of the DM substructure signal strength, we adopt a fiducial substructure model that follows the works by \citet{MASC2013}, assuming a moderate total boost factor of $b=33$ as given by their proposed parametrization of the boost for the Virgo mass (DM-I). We contrast this \emph{conservative} model with an \emph{optimistic} model that implicitly adopts a power-law extrapolation of the mass-concentration relation to the smallest (unresolved in simulations) halo masses, yielding $b=1200$ \citep[DM-II,][]{Gao2012b}.

For the spatial morphology of the DM-induced gamma-ray emission, including the predicted DM substructure signal, we adopt the form from a recent study of high resolution cosmological DM simulations of cluster-size halos \citep{Gao2012a}. The projected luminosity profile from the substructure is approximated by:
\begin{equation}\label{eq:sub}
  I_{\mathrm{sub}}(\theta) = \frac{16 b\times L_{\gamma}^{\mathrm{NFW}}}{\pi \ln(17)} \frac{1}{\theta_{200}^2+16 \theta^2},\quad \mathrm{for}\, \theta\le \theta_{200},
\end{equation}
where $\theta$ is the distance from the cluster center in degrees and $\theta_{200}$ is the angle subtending the virial radius, given by $\theta_{200}=\arctan(r_{200}/D_{a})\times 180\deg/\pi$, and $L_{\gamma}^{\mathrm{NFW}}$ is the total \gr luminosity of the halo within the virial radius \citep{Gao2012a}, defined as $L_{\gamma}=4\pi \times \Phi_{\gamma}\times D_{a}^{2}$.

\begin{figure}[tbp]
  \centering
  \includegraphics[width=.9\columnwidth]{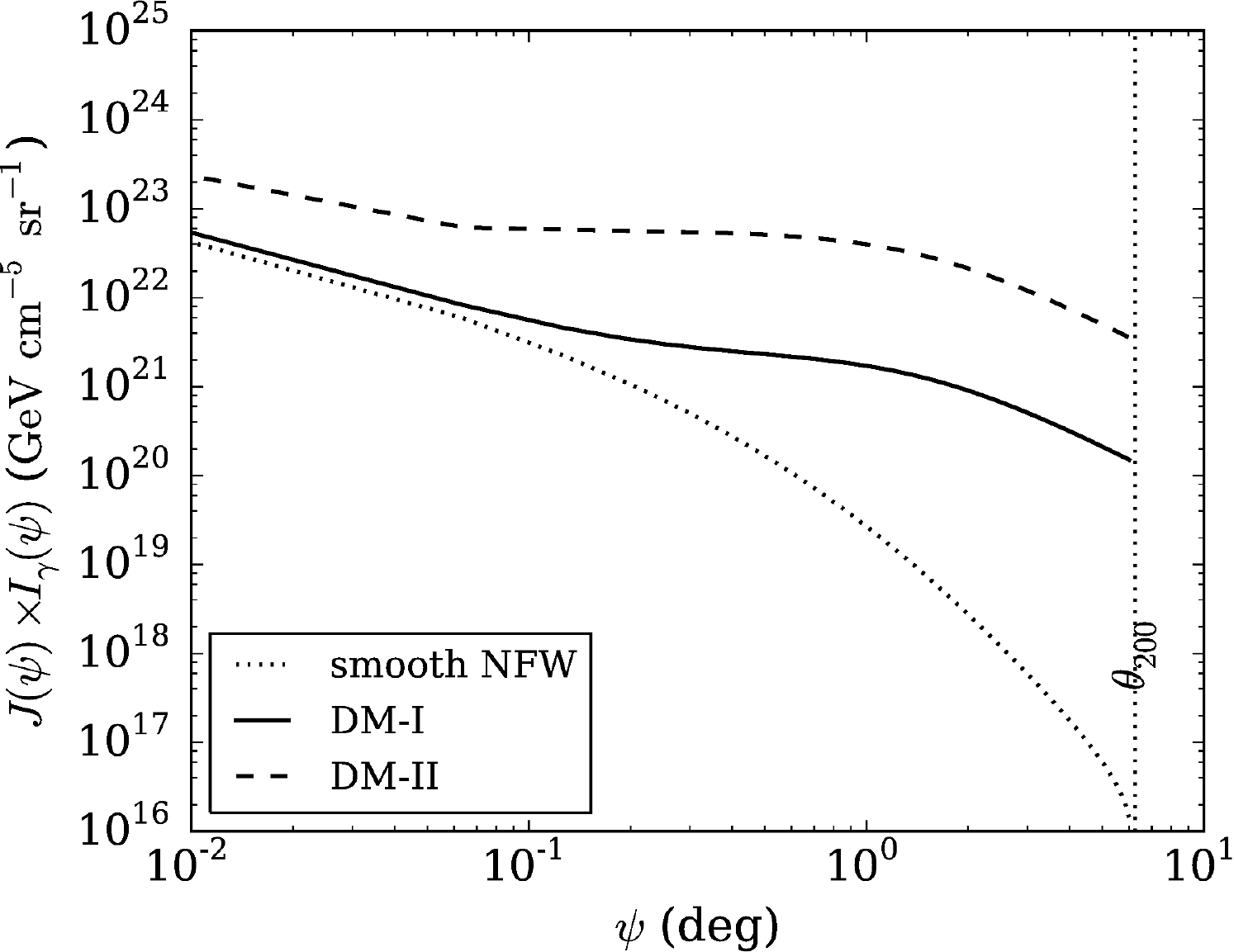}
  \caption{Shown is the annihilation flux profile as function of subtended angle for Virgo-I. We show this quantity for our two substructure benchmark scenarios (DM-I and DM-II) as solid and dashed lines, respectively. The dotted profile corresponds to the case where no substructure is included. For boosted profiles, the expected surface brightness profile has a broader (angular) distribution than for the smooth NFW profile. Outside the virial radius the DM halo is truncated, and accordingly we truncate our templates outside the virial radius. The dotted line indicates the angular virial radius $\theta_{200}$ (see text for details).\label{fig:profile_vs_r}}
\end{figure}

Note that the work by \citet{MASC2013} does not address the change in the spatial morphology of the annihilation signal due to the presence of sub-halos. Yet, in previous works, \citet{MASC2011} have shown that moderate values for {\it b} also lead to a significant flattening of the annihilation profile in clusters. We found this flattening to agree reasonably well with the one implied by Eq.~\ref{eq:sub} for moderate values of $b$. Thus, from here on, we assume this approximation to be a good representation of the spatial morphology of the DM substructure signal in both the DM-I and DM-II setups, with only the value of {\it b} differing from one to another substructure scenario. The resulting expected surface brightness profiles for DM annihilation are shown in Fig.~\ref{fig:profile_vs_r}.

Our choice of models for the substructure is motivated by assuming a common mass scale for both setups, $M_{\mathrm{cut}}=10^{-6}\msol$, at which the matter power-spectrum is truncated. Below this scale, no DM halos are formed and thus no halos will contribute to the expected DM signal. What the minimal DM halo size is and what properties these substructures have is, to a large extent uncertain and may depend on the specific DM particle model. The mass range for $M_{\mathrm{cut}}$ may vary from $10^7$ to $10^{-12}\msol$, where the upper limit comes the fact that we observe dwarf galaxies with that mass \citep{McConnachie:2012aa}. The lower limit is more uncertain and depends on the DM particle properties \citep{Profumo2006,Bringmann2009}.\footnote{Note that the actual values depend not only on the specific DM particle but also on the cosmological evolution of DM halos.}

\begin{figure}[tbp]
  \centering
  \includegraphics[width=.9\columnwidth]{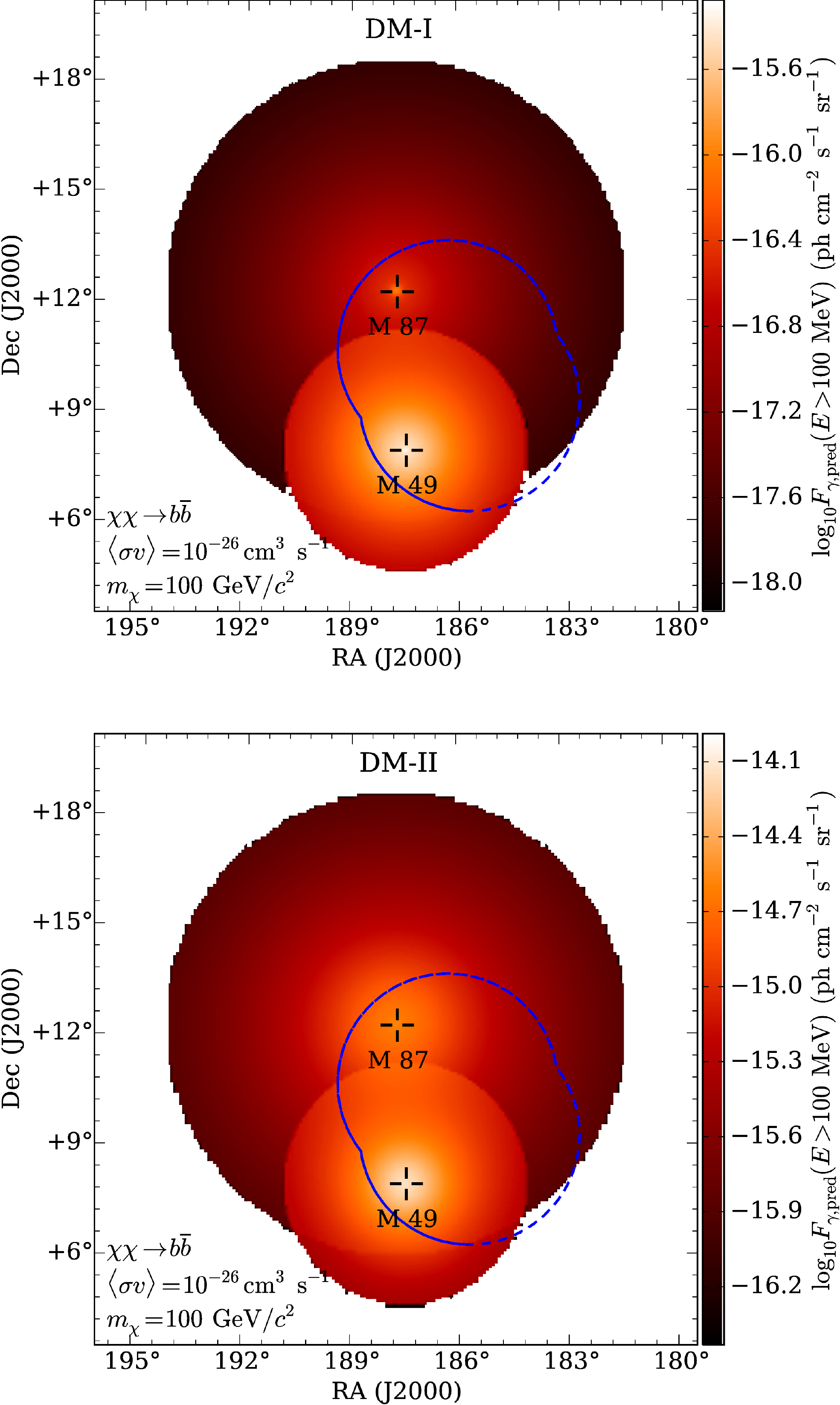}
  \caption{\label{fig:DMproj}Predicted integrated \gr-flux projection for the entire cluster above 100\,\mev for DM annihilation of a 100\,\gev WIMP into 100\% $b\overline{b}$ with an annihilation cross section of $10^{-26}\mathrm{cm^{3}s^{-1}}$ for the DM models discussed in this article (\emph{top}: DM-I, \emph{bottom}: DM-II). Note the different scales. The dashed contour indicates the location of the \gr excess reported in Section~\ref{sec:extended}.}
\end{figure}

Recalling the discussion on the morphology of the cluster from Section~\ref{sec:extended}, we remark that while the mass of Virgo-II is only about $\sim13\%$ of that of Virgo-I, its concentration parameter, $c=5.58$, is larger than that of Virgo-I (see Table~\ref{tab:jfactors}). {\edit{Since the DM annihilation flux is proportional to the third power of the concentration, the predicted DM-induced \gr flux of Virgo-II corresponds to about one third of the predicted DM-induced \gr flux from Virgo-I.}} In order to account for this in our DM modeling, we consider the cluster as a merging system where each sub-cluster is modeled individually according to the description given in this section. We then co-add the two resulting templates to form a composite spatial map which we use in the likelihood analysis. The projected annihilation flux maps are shown for the different models we choose in Fig.~\ref{fig:DMproj}. In this figure we added a contour that indicates the spatial position of the excess we report in this paper and we stress that a DM origin is unlikely because of the large offset between the predicted DM annihilation profile and the contours of the \gr excess (see Section~\ref{sec:extended} for details).

We summarize the main characteristics of our chosen models in Tab.~\ref{tab:jfactors}. We take the sum of the $J$-factors for Virgo-I and Virgo-II to be the total $J$-factor of the Virgo system.

\begin{deluxetable*}{lccccccc}
  \tablewidth{0pt}
  \tablecaption{Virgo subclusters and derived DM density profiles}

  \tablehead{
    \colhead{Sub-cluster} & \colhead{$M_{200}$} & \colhead{$r_{200}$} & \colhead{$\theta_{200}$} & \colhead{$c$\tablenotemark{a}} & \colhead{$J_{\mathrm{NFW}}$} & \colhead{$J_{\mathrm{DM-I}}$\tablenotemark{a}} & \colhead{$J_{\mathrm{DM-II}}$\tablenotemark{b}} \\
    \colhead{} & \colhead{($\times10^{14}\mathrm{M_{\odot}}$)} & \colhead{($\mathrm{Mpc}$)} & \colhead{($^\circ$)} & \colhead{} & \colhead{($\times10^{17}$)} & \colhead{($\times10^{18}$)} & \colhead{($\times10^{20}$)} }

  \startdata
  M87 (Virgo-I) & 5.60 & 1.70 & 6.3 & 4.21 & 2.56 & 6.50 & 3.33 \\
  M49 (Virgo-II) & 0.72\tablenotemark{c} & 0.88 & 3.8 & 5.58 & 1.85 & 5.36 & 0.75 \\
  \enddata
  \tablecomments{\label{tab:jfactors}Shown are the characteristic quantities used to derive the resulting $J$-factors for the Virgo cluster modeled as a merging system between the sub-clusters associated with M87 and M49. Columns from left to right are name, mass, virial radius, angular radius $\theta_{200}$, concentration parameter $c$, as well as $J$-factors for NFW and the DM models used in this analysis for each of the sub-clusters. All $J$-factors are given in units of $\mathrm{GeV^{2}cm^{-5}}$ and have been computed over a solid angle subtending the virial radius of each sub-cluster.}
  \tablenotetext{a}{\citet{MASC2013}}
  \tablenotetext{b}{\citet{Gao2012a}}
  \tablenotetext{c}{\citet{chen2007}}
\end{deluxetable*}

\subsection{DM Flux Enhancement due to Inverse Compton Scattering\label{sec:ICS}}
In calculating the $\gamma$-ray spectra from DM annihilation to leptons, we include the effects of Inverse Compton (IC) scattering of background radiation by electrons and positrons that result from the annihilation.\footnote{By leptons we refer to $e^{\pm}$ and $\mu^{\pm}$. We refrain from including IC calculations for annihilation into $\tau^{\pm}$ since its decay signature is closer to hadronic final states and thus any IC contribution would be sub-dominant.} We calculate the IC component of the spectrum by conservatively assuming scattering only of the cosmic microwave background (CMB); other radiation fields such as starlight could also contribute but are sub-dominant. We use the program DMFit \citep{Jeltema2008,Ackermann:2014aa} for spectrum calculations and include IC according to the procedure outlined in \citet{2010JCAP...05..025A}.

In cluster environments, electrons and positrons lose energy via radiation (e.g. IC scattering and synchrotron emission) on much shorter timescales than they diffuse. We therefore neglect the effects of diffusion \citep[e.g.][]{Colafrancesco2006}. We also neglect energy losses due to synchrotron radiation. Synchrotron losses would significantly suppress the IC signal if the average magnetic field of the cluster, $\langle B \rangle > B_{\mathrm{CMB}}\sim 3~\mu$G, where $B_{\mathrm{CMB}}$ is the magnetic field that has the same energy density of the CMB (the IC scattering background). Suppression would be on the scale of $(B_{\mathrm{CMB}}/ \langle B \rangle)^2 $. While data on the intracluster magnetic field of Virgo are limited, simulations suggest an averaged magnetic field of $\sim O(0.1\textrm{--}1~\mu\mathrm{G})$ \citep{Dolag2005a}, too small for synchrotron emission to be significant.

\subsection{Limits on \sigv}
We derive upper limits on \sigv using the profile likelihood method~\citep{Rolke2005} as implemented in the {\tt MINOS}-subroutine of the {\tt MINUIT} package \citep{James1975} which is available through the \emph{Fermi} Science Tools. We define the 95~\% upper limit on \sigv as the value of \sigv for a given mass $m_\chi$ where twice the difference in the \LL, $2\times\Delta\mathcal{L}=2.71$ with respect to the value of the \LL\,  for the best fit value.\footnote{For the background modeling we employ the same considerations as discussed in Section~\ref{sec:ana}.}

\begin{figure*}[htbp]
  \centering
  \includegraphics[width=.9\columnwidth]{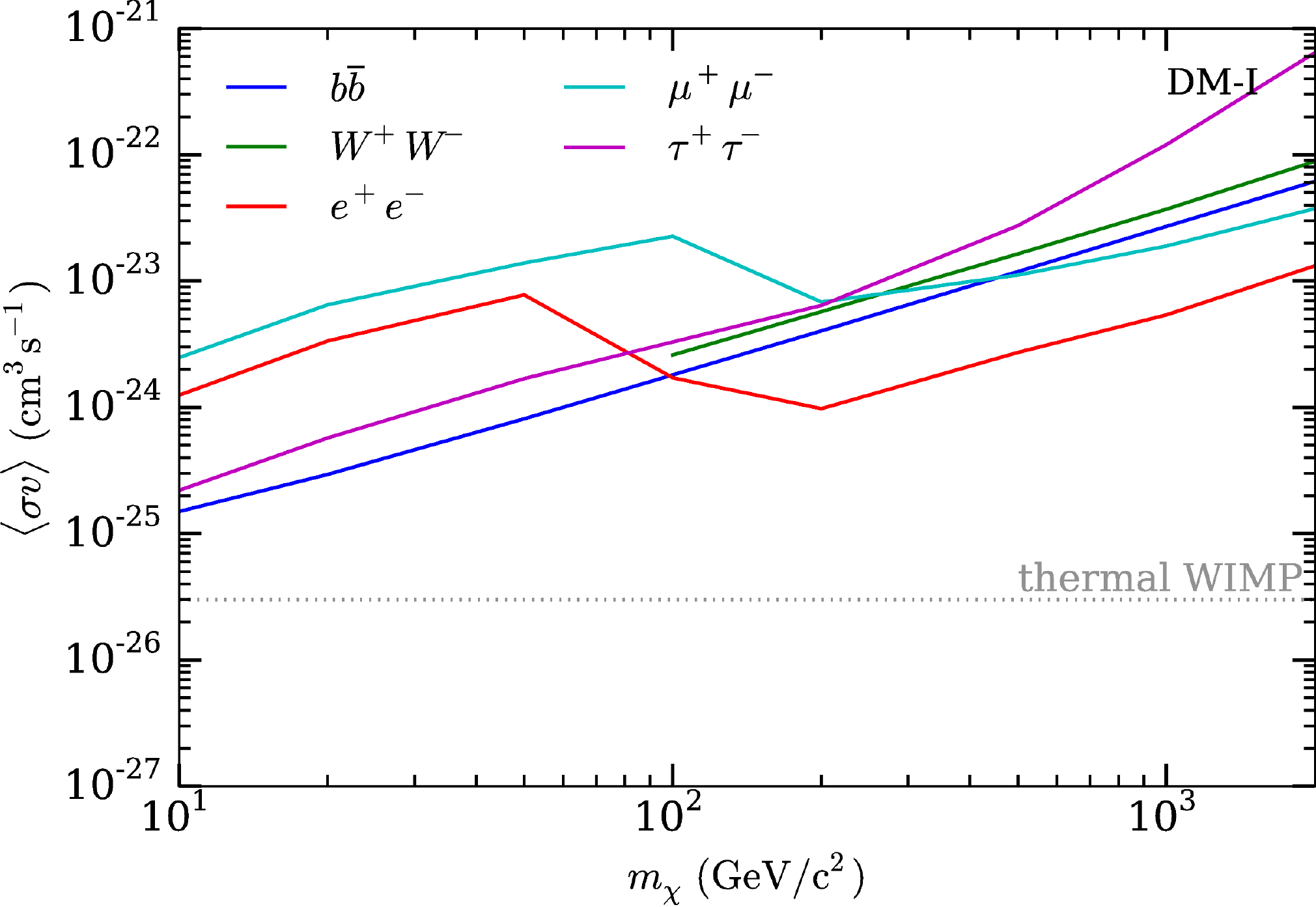}
  \includegraphics[width=.9\columnwidth]{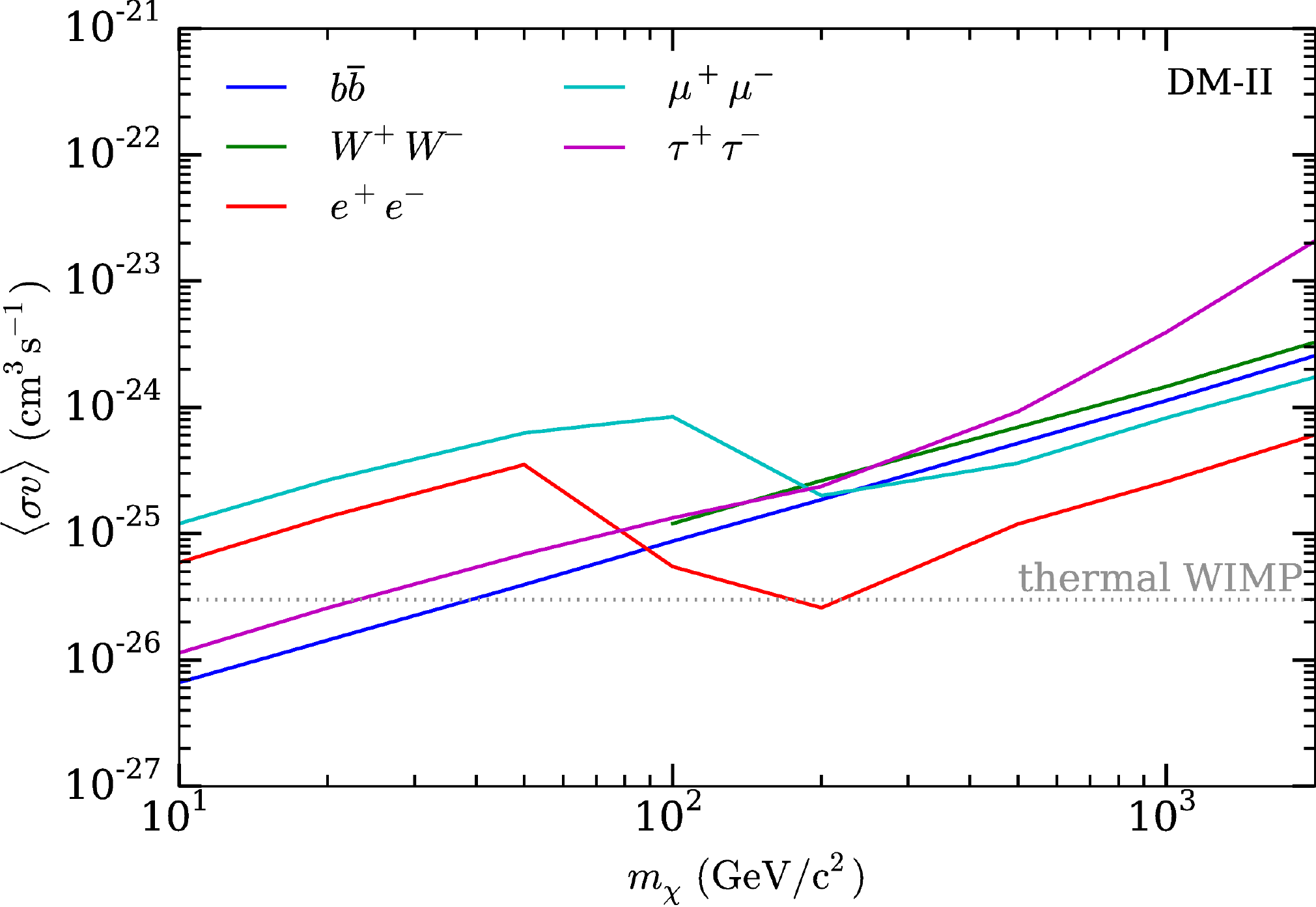}
  \caption{\label{fig:channels}Obtained 95\% CL upper limit on \sigv for various annihilation channels assuming our fiducial substructure models (\emph{top}: DM-I, \emph{bottom}: DM-II). Both $e^{\pm}$ and $\mu^{\pm}$ channels include the contribution from IC scattering with the CMB as detailed in Section~\ref{sec:ICS} which starts to dominate the predicted emission above $50\,\gev$ for $e^{\pm}$ and $100\,\gev$ for $\mu^{\pm}$. The dashed line corresponds to the annihilation cross section for a thermal WIMP.}
\end{figure*}

Fig.~\ref{fig:channels} shows the dependency of the upper limits on the chosen DM annihilation channel for our fiducial models. The most constrained channels are $\chi \chi \rightarrow b\overline{b}$ and $\chi \chi \rightarrow \tau^+ \tau^-$. Accounting for IC emission in the leptonic channels $e^\pm$ and $\mu^\pm$ improves the constraints we obtain from the prompt emission by two to three orders of magnitude, above DM masses of $50\,\gev$ and $100\,\gev$, respectively. The limits for $e^\pm$ are the most constraining for DM masses above $\sim110\,\gev$, due to the enhanced flux predictions from IC.

\begin{figure}[htbp]
  \centering
  \includegraphics[width=.9\columnwidth]{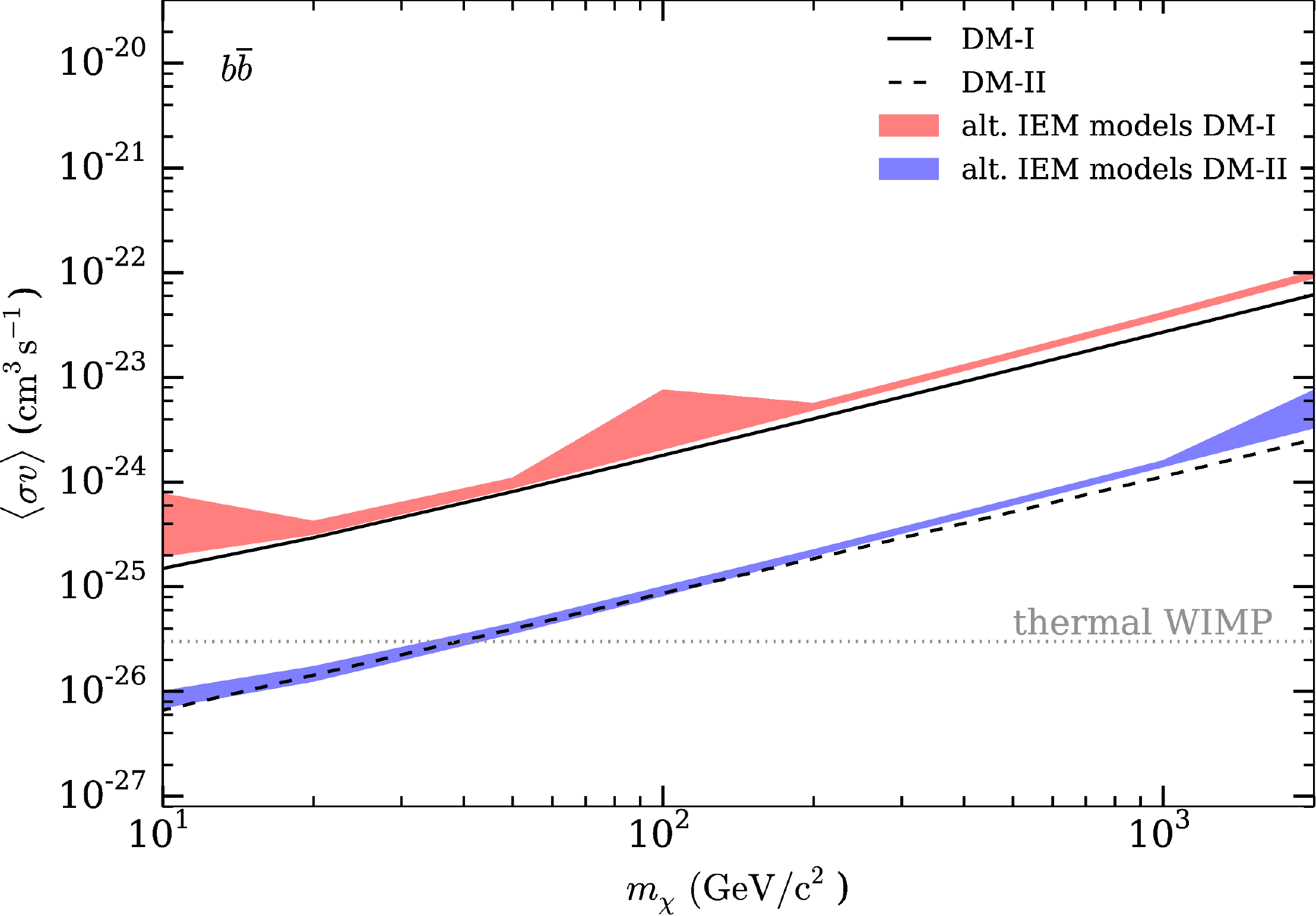}\\
  \includegraphics[width=.88\columnwidth]{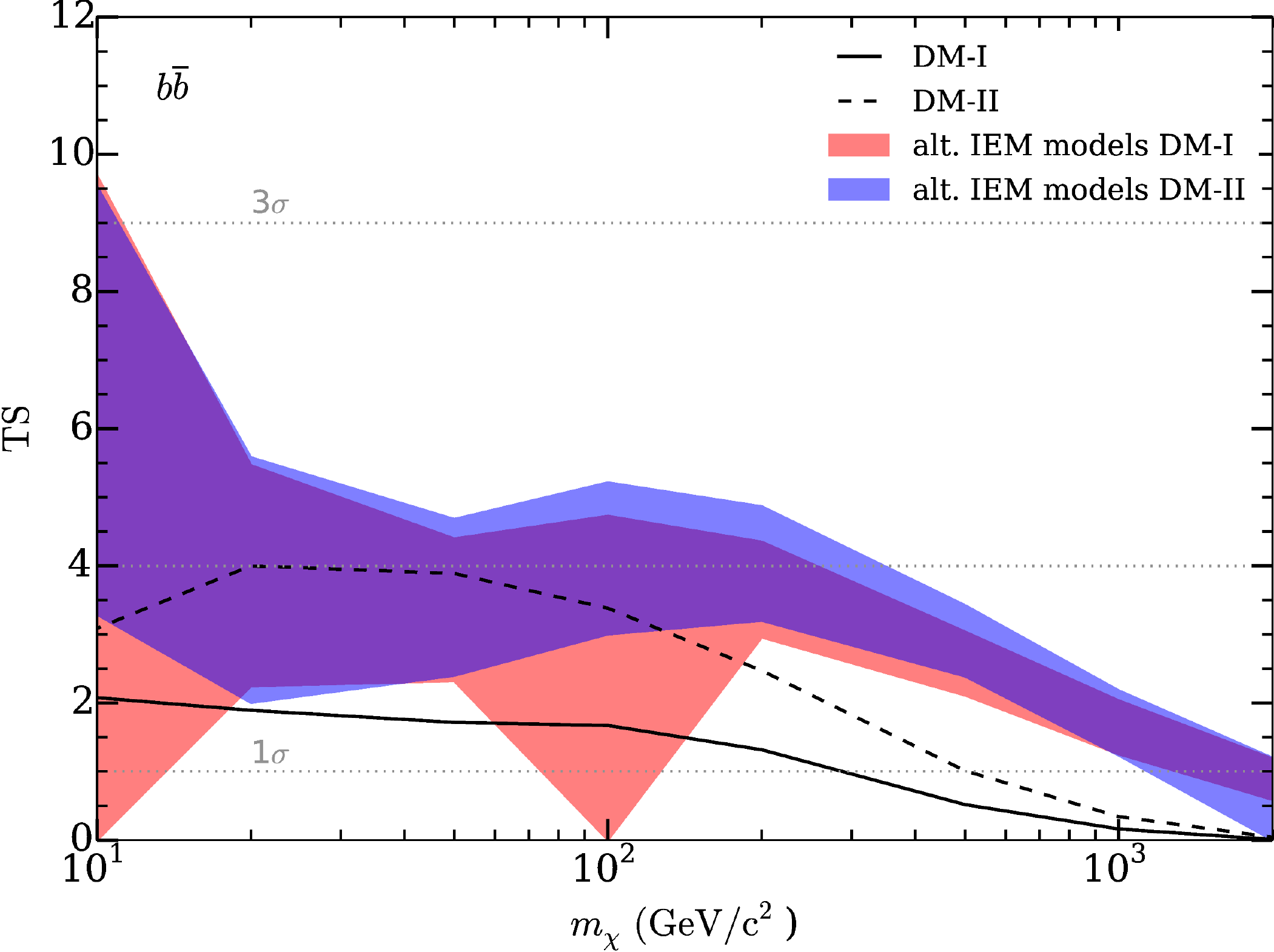}
  \caption{ \emph{Top}: Obtained 95\% CL upper limit on \sigv for a DM WIMP annihilating into $b\overline{b}$ in the mass range from 10~\gev up to 2~\tev. The shaded areas represent the range of limits obtained when replacing the standard \iem with the alternative models described in Section~\ref{sec:IEM}. Solid and dashed lines represent the limits obtained using the standard \iem for our conservative (DM-I) and our optimistic (DM-II) boost model, respectively. The dashed line corresponds to the annihilation cross section of a thermal WIMP. \emph{Bottom}: Shown are the associated $TS$ values with this choice of models. See the text for a discussion regarding the $TS$-values obtained with the alternative diffuse models. Note that in both plots we omit data points in which Minuit/MINOS did not reach convergence ($<10\%$ of the tested mass-model scan points).}
  \label{fig:dm_diffuse}
\end{figure}

In Fig.~\ref{fig:dm_diffuse} we show our derived upper limits on \sigv and their associated $TS$ values for the $\chi \chi \rightarrow b\overline{b}$ channel and contrast our standard \iem with results obtained from using the alternative diffuse models as discussed in Section~\ref{sec:IEM}. Our optimistic limits exclude thermal WIMP cross sections below $40\,\gev$. The limits derived from the more conservative assumptions are a factor 20 weaker across the entire probed mass range. Even with the inclusion of additional point sources as done in this work, there is a residual $TS\sim4$ if we consider the more extended and elongated profile as predicted by our optimistic model (DM-II). For the DM-I case, this value is reduced even further. Considering the alternative diffuse models, the resulting limits are generally weaker, associated with residual $TS<5$ except for WIMP masses $\lesssim20\,\gev$. For lower masses, the alternative models give rise to residual $TS$ peaking at $TS\sim9.5$ or $\sim3.1\sigma$. Low-mass DM models associated to relatively high $TS$-values for one diffuse model show a large spread ($\delta TS\simeq5$) in $TS$ for the alternative models.

Recalling the description of the alternative diffuse models in Section~\ref{sec:IEM}, these differ from the standard \iem\, by having various large-scale components fit freely to the data (e.g. Loop~I, IC, etc.). The extent of these large-scale components is comparable to the spatial extension of our cluster template which causes a degeneracy between the fit parameters for the diffuse components and Virgo. As a consequence we find that soft photons ($E\lesssim10\,\gev$), which would otherwise be attributed to the background \iem\, are now included in the number of predicted photon counts from Virgo for a light WIMP model.\footnote{For illustration purposes, the reader is reminded that the typical \gr spectrum (energy flux) of, e.g., a 20~\gev WIMP annihilating into $b\overline{b}$ peaks at $\sim2\,\gev$ and results mainly in soft photons in the \mev-\gev range, which can explain the large spread towards the lowest WIMP masses shown in Fig~\ref{fig:dm_diffuse}.} Note that this effect appears to be even more pronounced as the spatial template for the Virgo cluster is even more extended than the disk used in our previous study (refer to Section~\ref{sec:extended} for a detailed discussion). Finally, we also remark that this issue is by construction less apparent for the standard \iem, since here all components are fixed to their relative best-fit contributions obtained from a likelihood fit to the entire \gr-sky.

\section{Cosmic-Ray-Induced Gamma Rays}\label{sec:crVirgo}
An alternative production mechanism of \grs originating from the Virgo region may be due to CR interactions. \grs are mainly produced in IC interactions of relativistic electrons or via hadronic $pp$-collisions producing pions and \grs through $\pi^{0}\rightarrow2\gamma$ \citep{Brunetti:2012aa}. The dominant production mechanism of \gr\ from CRs in the ICM is still debated: either cosmic-rays are accelerated directly in structure formation shocks (including the effect of AGNs and supernovae) through diffusive shock acceleration or an aged population of cosmic-ray are reaccelerated in the turbulent plasma generated by e.g. merging clusters \citep[see, e.g.][for a review]{2014IJMPD..2330007B}.

Since there is no giant radio halo associated with the Virgo cluster and the central part of the cluster has properties similar to a cool-core cluster \citep{2011MNRAS.414.2101U}, we expect the \grs from a population of reaccelerated cosmic-rays \citep[see, e.g.][]{2013ApJ...762...69Z} to be too faint to be detectable by the \lat\ throughout its lifetime. However, there is a strong dependence on the uncertain turbulent profile. Indeed, \citet{Pinzke:2015aa} showed that for a flatter turbulent profile than what was previously assumed, the \gr emission could be in reach with Ferm-LAT in the coming years. To keep the CR analysis simple, we neglect these aforementioned models as well as other leptonic models \citep{2009JCAP...08..002K}. Instead, we focus on constraining the \grs produced in a pure hadronic scenario in that region. Specifically, we adopt a simple but realistic model for the predicted universality of the CR-spectra built up from diffusive shock acceleration in large-scale structure formation shocks \citep{2008MNRAS.385.1242P,2010MNRAS.409..449P}. Based on these considerations, in this section we derive constraints on the CR-induced \gr\ flux and related CR quantities from Virgo.

\subsection{Modeling and Results}
\begin{figure*}
  \begin{center}
    \includegraphics[width=.9\textwidth]{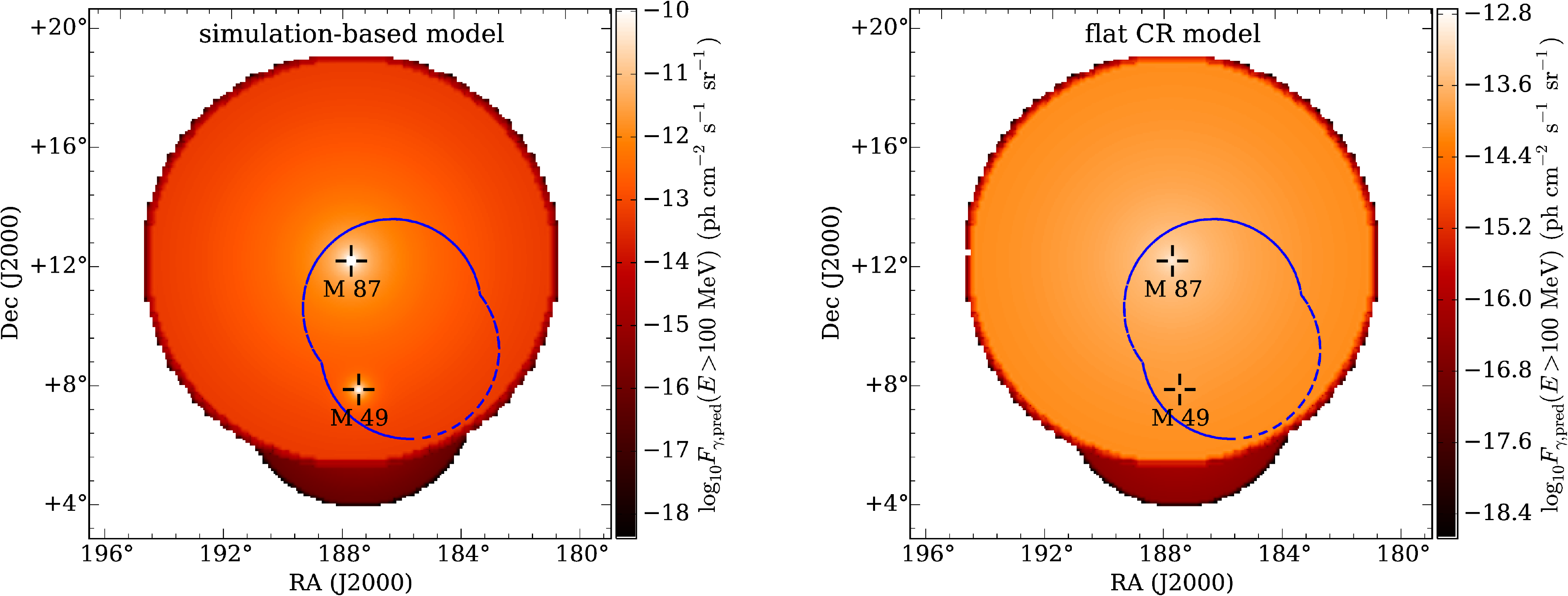}
  \end{center}
  \caption{Projected predicted, integrated CR-induced \gr\ flux (above $E=100\,\mev$) for the models considered in this analysis (\emph{left:} simulation-based model following \citet{Pinzke2011}; \emph{right:} model in which the CRs follow a flat distribution)  in units of $\mathrm{ph\,cm^{-2}\,s^{-1}\,sr^{-1}}$. Each model is a superposition of the individual CR-models derived for M87 and M49. For reference, we show the location of the excess as blue dashed contour. Note the different scales in both plots. \label{fig:CRmorph}}
\end{figure*}

Following earlier works in \citet{Ackermann2013}, we consider two different (hadronic) models for the CR distribution, the simulation-based approach by \citet{Pinzke2011}, which predicts a \gr surface brightness which closely follows the X-ray emitting gas in the ICM, and a model in which the CRs are confined within the cluster virial radius but evenly distributed with no dependence on the ICM gas (flat model). The latter can thus be seen as a simplified proxy for CR-streaming models which can lead to more extended \gr brightness profiles \citep{2011A&A...527A..99E, 2013MNRAS.434.2209W, 2014MNRAS.440..663Z}. While the expected \gr morphology varies, we assume the spectrum to be approximated by the universal  model as detailed in \citet{Pinzke2011} (the interested reader is referred to Figure~1 of that paper). Analogously with the results presented in Section~\ref{sec:DMsubstructure}, we construct a model which takes into account the merging state of the cluster by overlaying the spatial template inferred from X-ray profiles from Virgo-I with that of Virgo-II. We show the predicted flux maps in Fig.~\ref{fig:CRmorph}. Outside $r_{200}$ we take the predicted flux to be negligible.

In analogy with the results in the previous section, we use the profile likelihood method to derive 95\% upper limits on the CR-induced \gr\ flux. Our results
are shown in Table~\ref{tab:CRresults}.

\begin{deluxetable*}{ccccc}
  \tablecaption{CR-models and derived limits}
  \tablehead{\colhead{CR model} & \colhead{$F_{\gamma,\mathrm{pred}}(E>100\,\mev)$} & \colhead{$F_{\gamma,95}(E>100\,\mev)$} & \colhead{$\langle X_{\mathrm{CR}} \rangle$} & \colhead{\zetap}\\
    \colhead{} & \colhead{$(\times 10^{-9}\mathrm{ph\,cm^{-2}\,s^{-1}})$} & \colhead{$(\times 10^{-8}\mathrm{ph\,cm^{-2}\,s^{-1}})$} & \colhead{} & \colhead{}
  }
  \startdata
  Simulation-based & 15.0 & 1.2 & 6\%&40\%\\
  Flat CR\tablenotemark{a} &0.4 & 1.8 &\dots & \dots \\
  \enddata

  \tablecomments{\label{tab:CRresults}Shown are both predicted and observed integrated fluxes above 100~\mev for our models for CR-induced \grs\ as discussed in the text. For the simulation-based model, the remaining columns denote the volume-averaged CR-to-thermal pressure ratio and the maximum acceleration efficiency for CR protons, respectively.\tablenotemark{b}}
  \tablenotetext{a}{In order to provide a consistent description, we normalize each profile to the total CR number within $r_{200}$ for the simulation-based model.}
  \tablenotetext{b}{As the observed flux for the flat CR-model is a factor $\sim45$ above the predictions, these limits cannot be used to constrain \xcr.}
\end{deluxetable*}

We exclude \gr integral fluxes above $1.2\times10^{-8}\mathrm{ph\,cm^{-2}\,s^{-1}}$ for the simulation-based CR model over the energy range 100~\mev -- 100~\gev, which is about a factor $\sim1.4$ stronger than previously published \citep{2010ApJ...717L..71A}. Using the flat model yields an integral flux limit of $1.8\times10^{-8}\mathrm{ph\,cm^{-2}\,s^{-1}}$ which is above the value that was published previously. This can however be explained by the fact that flat CR models are generically less constrained by current \gr\ data \citep{Ackermann2013,2014MNRAS.440..663Z}.

\subsection{Constraints on \zetap and \xcr}
Two important quantities associated with CRs are the maximum efficiency with which CRs are accelerated in shocks, \zetap, along with the volume-averaged CR-to-thermal pressure ratio, \xcr. Current limits exclude efficiencies above 21\% and values for $\xcr>1\%$ for purely hadronic models \citep{Ackermann2013,2014MNRAS.440..663Z}. As shown in \citet{Ackermann2013}, for the simulation-based CR model, we expect a linear relationship between the \gr flux (or the limit on the flux) and \xcr as well as \zetap, respectively,  with little variation across cluster masses and evolutionary stages \citep{Pinzke2011}.

As the additional point sources are not fully sufficient to model the entirety of the reported \gr\ excess, the resulting limits from CR physics are less constrained. We find $\zetap\leq40\%$ and $\xcr\leq6\%$ within $r_{200}$ of the combined system, both of which ranges have been excluded in previous multi-sample studies \citep{Ackermann2013} (all limits have been derived at a 95\% confidence level).

\subsection{Systematic Uncertainties due to \iem modeling}
In order to assess the robustness of these results, we repeat the calculations in the previous section for our set of alternative diffuse models. We find that our derived constraints can be up to $\sim40\%$ better than those obtained with the standard \iem.

\subsection{Degeneracy of Results with M87}
In general, CR-induced models are substantially more centrally peaked than any of our previously considered DM-motivated models \citep[see, e.g.][for a study of various CR scenarios, in the Coma cluster]{2014MNRAS.440..663Z}. This implies the potential for degeneracy with M87 itself (now referring to the AGN and not to the sub-cluster). Detected with the \lat\ with only six months exposure, M87 \citep{2009ApJ...707...55A} is best modeled as a power law with $\Gamma=2.1$ which is harder than the tested CR-models (above $\sim1\,\gev$ the CR-model by \citet{Pinzke2011} can be approximated as a power-law with $\Gamma=2.3$). When comparing the fit results of the spectral parameters of M87 (both index and normalization are left to vary freely in the fit), we find that these vary within the quoted uncertainty given in 2FGL when performing the likelihood fit including either CR-model discussed here. We also note that since the cluster is modeled as merging system rather than as a spherically symmetric object, this helps in breaking the degeneracy between M87 and any cluster-induced emission.

\section{Conclusion}\label{sec:conclusion}
We find no strong evidence for extended emission associated with the Virgo cluster center. Yet, using the standard \iem we find a statistically significant extended excess from a disk profile with radius 3\deg clearly offset from the cluster center. Our TS map reveals two well-separated maxima, both clearly offset from the two main sub-clusters associated with the giant ellipticals M87 and M49. This signature makes a DM origin unlikely. Also, as there is no indication of accelerated CRs, evidenced by either radio or X-ray emission, an astrophysical origin due to e.g. accelerated CRs in the ICM is questionable. We thus report upper limits on CR-scenarios and DM-induced \grs.

Similar to previous studies, we carry out a search for new point sources in order to account for the increased data volume with respect to the employed source catalog. We find six new candidates in accordance with similar studies by~\citet{Macias} and \citet{HanII}. These new candidates, however, have no reported counterparts in other wavebands. Five of them are contained in the 3FGL-catalog. We carry out an alternative \iem study which is essential for estimating systematic uncertainties associated with the search for \gr\ emission from very extended sources. In our case the inconsistency between the \iems  demonstrates that the Virgo region is an especially difficult section of the sky. The proximity to poorly understood Galactic foregrounds emitting \grs, like Loop~I, makes the search for extended emission from this region very challenging. Our study also reveals the challenges of searches for such low photon density sources even in at high Galactic latitudes.

Accounting for the complex dynamics of the cluster, we model its emission by co-adding the contributions from the major sub-clusters centered on M87 and M49, respectively. In particular for very extended models, as predicted if considering large amounts of DM halo substructure, the spatial morphology departs from spherical symmetry. Resulting limits for either DM- or CR-induced \grs are generally weaker than that of other targets, e.g. dwarf spheroidal for the case of DM-annihilation \citep{dwarf_paper} and from collective cluster studies \citep{Ackermann2013}. The DM limits from the Virgo analysis here, for instance, are about an order of magnitude above the thermal WIMP cross-section when assuming a realistic model for the sub-halo boost.

Finally we would like to stress that the main findings in this paper are expected to remain unchanged even if more data were to be included, as the uncertainties in the results are dominated by systematics associated with the \iem modeling. We emphasize that the improved source model used with the analysis roughly corresponds to the model presented in the current deepest \gr\ catalog, 3FGL. Also, while the predicted constraints on DM annihilation and CR processes improve by up to a factor of $\simeq1.4$ if all available data are considered (6 years instead of 3 years), targets other than the Virgo cluster may be better suited for analysis, e.g. the Coma cluster for CR processes and Fornax for DM prospects \citep[see, e.g.][for a discussion]{Pinzke2011,2012JCAP...07..017A}.\footnote{\edit{Note that recent work by \citet{Feyereisen:2015aa} suggests that a detection of DM-induced \grs from clusters is unlikely, especially in the presence of large substructure boosts, since the same signature constitutes an irreducible background to the isotropic \gr background \citep{Ackermann:2015ab}.}} While farther away, the predicted \gr emission from both clusters is expected to be within the detection reach of the LAT and their apparent extensions on the sky is significantly less than Virgo which helps to reduce the uncertainties associated with the foreground \iem modeling, thus allowing for a more robust analysis.

\section*{Acknowledgments}
AP is grateful to the Swedish Research Council for financial support.

The \textit{Fermi} LAT Collaboration acknowledges generous ongoing support from a number of agencies and institutes that have supported both the development and the operation of the LAT as well as scientific data analysis. These include the National Aeronautics and Space Administration and the Department of Energy in the United States, the Commissariat \`a l'Energie Atomique and the Centre National de la Recherche Scientifique / Institut National de Physique Nucl\'eaire et de Physique des Particules in France, the Agenzia Spaziale Italiana and the Istituto Nazionale di Fisica Nucleare in Italy, the Ministry of Education, Culture, Sports, Science and Technology (MEXT), High Energy Accelerator Research Organization (KEK) and Japan Aerospace Exploration Agency (JAXA) in Japan, and the K.~A.~Wallenberg Foundation, the Swedish Research Council and the Swedish National Space Board in Sweden.

Additional support for science analysis during the operations phase is gratefully acknowledged from the Istituto Nazionale di Astrofisica in Italy and the Centre National d'\'Etudes Spatiales in France.

Facilities: \facility{\emph{Fermi}}


\end{document}